\documentclass[fleqn,usenatbib]{mnras}
\usepackage{newtxtext,newtxmath}
\usepackage[T1]{fontenc}
\usepackage{ae,aecompl}

\usepackage[version=4]{mhchem}    % Chemistry stuff
\usepackage{graphicx}	% Including figure files
\usepackage{amsmath}	% Advanced maths commands
\usepackage{amssymb}	% Extra maths symbols
\title[Nitrogen dominated atmospheres of hot super-Earths]{Atmospheric compositions and observability of nitrogen dominated ultra-short period super-Earths}

\author[M.Zilinskas et al.]{
Mantas Zilinskas,$^{1}$\thanks{E-mail: zilinskas@strw.leidenuniv.nl}
Yamila Miguel$^{1}$
Paul Molli\`ere$^{2}$
Shang-Min Tsai$^{3}$
\\
$^{1}$Leiden Observatory, Leiden University, Niels Bohrweg 2, 2333CA Leiden, The Netherlands\\
$^{2}$Max-Planck-Institut f\"{u}r Astronomie , K\"{o}nigstuhl 17, 69117 Heidelberg, Germany\\
$^{3}$Atmospheric, Ocean, and Planetary Physics, Department of Physics, Oxford University, OX1 3PU, United Kingdom
\\
}

\date{Accepted 2020 March 7. Received 2020 March 2; in original form 2020 January 9.}

\pubyear{2019}

\begin{document}
\label{firstpage}
\pagerange{\pageref{firstpage}--\pageref{lastpage}}
\maketitle

\begin{abstract}
We explore the chemistry and observability of nitrogen dominated atmospheres for ultra-short-period super-Earths. We base the assumption, that super-Earths could have nitrogen filled atmospheres, on observations of 55 Cnc e that favour a scenario with a high-mean-molecular-weight atmosphere. We take Titan's elemental budget as our starting point and using chemical kinetics compute a large range of possible compositions for a hot super-Earth. We use analytical temperature profiles and explore a parameter space spanning orders of magnitude in C/O \& N/O ratios, while always keeping nitrogen the dominant component. We generate synthetic transmission and emission spectra and assess their potential observability with the future James Webb Space Telescope and ARIEL. Our results suggest that \ce{HCN} is a strong indicator of a high C/O ratio, which is similar to what is found for H-dominated atmospheres. We find that these worlds are likely to possess C/O > 1.0, and that \ce{HCN}, \ce{CN}, \ce{CO} should be the primary molecules to be searched for in thermal emission. For lower temperatures (T < 1500 K), we additionally find \ce{NH3} in high N/O ratio cases, and \ce{C2H4}, \ce{CH4} in low N/O ratio cases to be strong absorbers. Depletion of hydrogen in such atmospheres would make \ce{CN}, \ce{CO} and \ce{NO} exceptionally prominent molecules to look for in the 0.6 - 5.0 $\micron$ range. Our models show that the upcoming JWST and ARIEL missions will be able to distinguish atmospheric compositions of ultra-short period super-Earths with unprecedented confidence.
\end{abstract}


%%%%%%%%%%%%%%%%%%%%%%%%%%%%%%%%%%%%%%%%%%%%%%%%%%

%%%%%%%%%%%%%%%%% BODY OF PAPER %%%%%%%%%%%%%%%%%%

\section{Introduction}

%Until just recently, the entire nature of ``super-Earths", planets that are larger than Earth but smaller than Neptune, has been largely shrouded in mystery. 
Until recently, the elusive nature of ``super-Earths" has been largely shrouded in mystery. Population studies and characterisation efforts have shown that these planets are amongst the most occurring around Sun-like and M-dwarf stars \citep{Petigura_2013,Dressing_2015,Burke_2015}. It was later deduced that the observed population exhibits a bimodal distribution in radii \citep[and the references therein]{Fulton_2017,Fulton_2018} that is thought to be caused by photoevaporation. The process divides the observed population into predominantly rocky, high-density super-Earths and to super-Earths with thick, hydrogen-rich atmospheres, also known as sub-Neptunes. In this paper we focus on the first category, consisting of super-Earths, which we take to be up to 2.0 R$_{\earth}$, with ultra-short periods (USP) and atmospheres that make minimal contribution to their total size. Some of these super-Earths might have metal-rich atmospheres outgassed from the lava oceans \citep{Schaefer_2010,Miguel_2011}, while others might possess a heavier atmosphere \citep{Demory_2016b}.

%Some of these super-Earths, which we take to be up to 2.0 R$_{\earth}$, orbit their star at distances, where the irradiation temperature is sufficient to melt and vaporise surface rock, allowing for outgassed, low-pressure and metal-rich atmospheres \citet{Miguel_2011} (\textit{Note sure if need this sentence}).

With a period of $P = 0.7365474 \pm 1.3 \times 10^{-6}$ days (orbital distance of 0.0154 AU) and radius of $1.947 \pm 0.038$ $R_{\earth}$ \citep{Bourrier_2018,Crida_2018}, 55 Cancri e (55 Cnc e) is an extreme example of such a world. Planets in USP orbits are tidally locked and extremely hot on their day side, which makes them particularly good targets for thermal emission observations. 55 Cnc e itself has been observed a multitude of times \citep{demory_2011,demory_2012,Ehrenreich_2012,de_mooij_2014,Demory_2016a,Demory_2016b,Harper_2016,Tsiaras_2016a,Esteves_2017}. Despite the efforts, its atmospheric structure and composition still remain unknown.

Phase curve observations of 55 Cnc e by \citet{Demory_2016b} show a significant hot-spot shift which presents a case for a substantial atmosphere. Follow up analysis of the same phase curve data by \citet{Kite_2016,Angelo_2017} and GCM modelling by \citet{Hammond_2017} also strongly imply that the seen effectiveness of the advective heat redistribution on the planet cannot be explained solely via the surface lava currents, instead suggesting that an optically thick atmosphere is needed.

While the analysis of spectroscopic HST/WFC3 observations by \citet{Tsiaras_2016a} suggest that the atmosphere might be light-weighted, retaining a significant amount of hydrogen and helium, the interior structure modelling by \citet{Bourrier_2018} excludes any possibility of an extended H/He gas envelope.
This assertion is supported by the non-detection of \ce{H} or \ce{H2O} in the atmosphere by \citet{Ehrenreich_2012, Esteves_2017}. It is also expected that if 55 Cnc e were to possess a water-rich atmosphere, it would be unlikely to have had survived the intense erosion from its host star over its lifetime ($\sim$10 Gy) \citep{Bourrier_2018}. Although there have been efforts to detect molecular features of \ce{Na} and \ce{HCN} \citep{Harper_2016,Tsiaras_2016a} in the atmosphere of 55 Cnc e, the results were inconclusive. The summarised evidence leans towards indicating that the planet does possess a high-mean-molecular-weight atmosphere, but its exact composition is still unknown. Such an atmosphere could have \ce{CO} or \ce{N2} as its main component and is most likely up to a few bars in photospheric pressure \citep{Angelo_2017}.

\citet{Miguel_2019} has explored the possibility of a nitrogen based atmosphere using equilibrium chemistry and have found that the atmosphere might show spectral features of \ce{NH3}, \ce{HCN} and \ce{CO}. The purpose of this paper is to significantly expand upon those results and explore nitrogen based atmospheric compositions for planets representative of 55 Cnc e using chemical kinetics. Chemical kinetics allows us to incorporate, otherwise not present, photochemistry, molecular diffusion and vertical transport, all of which can have a substantial impact on the molecular composition of the observable parts of a planetary atmosphere. %\citep{Hu_2014,Tsai_2017}. 

We take Titan's elemental composition as our starting point and model a wide range of atmospheres with C/O and N/O ratios spanning multiple orders of magnitude. We do this for the maximum and minimum hemisphere-averaged temperatures of 55 Cnc e from \citet{Angelo_2017}, and take the analytical thermal profile approach as in \citet{Miguel_2019,Morley_2017}.

Using our computed molecular compositions, we present a range of synthetically generated spectra for 55 Cnc e, in both, transmission and thermal emission, covering the range from 0.6 to 28 $\mu$m. We show the potential observability of spectral features in the atmospheres of super-Earths similar to 55 Cnc e (with e.g. JWST and ARIEL) and identify key indicators of N-dominated atmospheres in the spectra of super-Earths. 

This paper is arranged as follows. In Section \ref{sec:methods} we describe our methodology, outlining the chemical, radiative transfer and JWST noise models. We present the resulting atmospheric abundances and the generated synthetic spectra in Section \ref{sec:results}. We interpret our key findings in Section \ref{sec:discussion}, with emphasis on variation of the most relevant atmospheric species with different C/O \& N/O ratios and their future observability. Section \ref{sec:conclusion} contains the summary and the conclusion of our study.

\section{Methods}
\label{sec:methods}
\subsection{Model Parameters}
\subsubsection{Atmospheric Compositions}
\label{sec:meth_atmos_comp}
The composition of a super-Earth's atmosphere heavily depends on its accretion history and geological activity. Unlike gaseous planets, these may not necessarily possess or retain a primary, hydrogen dominated atmosphere captured from the nebula during the formation process. Its small mass, evolution and the proximity to the host star are likely to drive the atmosphere to lose its light, volatile elements and become enriched with heavier species from the surface degassing or lava vaporisation. For hot, USP super-Earths, it is believed that the close proximity to the star will ultimately lead to escape of most of its H/He \citep{Lammer_2009,Lopez_2013,Owen_2013,Jin_2014}. 

Atmospheric compositions originating from the degassing of the accreted material heavily depend on which group the accreted material belongs to. Accretion of large amounts of certain carbonaceous chondrites, coming from the outer disk, could result in a very carbon-rich atmosphere, with C/O larger than unity. Accretion of enstatite or ordinary chondrites, from the inner part of the disk, would produce an atmosphere with C/O near unity \citep{Elkins_Tanton_2008,Schaefer_2010,Hu_2014}.

In general, the lack of observational evidence of super-Earth atmospheres and the large uncertainties in the formation and evolution processes present a possibility for the C/O ratios of these atmospheres to have significant variation from that of solar (C/O = 0.5) \citep{Hu_2014}. It is not inconceivable for the C/O ratio to be much higher than that of unity. With this context, we take the freedom to explore an unusual and diverse range of nitrogen dominated atmospheric compositions, spanning orders of magnitudes in C/O and N/O ratios. As in \citet{Morley_2017,Miguel_2019}, we use Titan's atmospheric composition of N = 0.96287, H = 0.03, C = 0.007 and O = 2.5 $\times 10^{-5}$ --in units of mass fractions-- as our starting point. While keeping the abundance of hydrogen constant, we simultaneously vary nitrogen, carbon and oxygen abundances by solving a system of linear equations with unique solutions. This effectively probes a parameter space encompassing C/O and N/O ratios important for the formation of relevant nitrogen and carbon species, e.g. \ce{HCN}. The C/O values used in this study range from 0.01 to 280. Additionally, for each C/O value we also explore the effects of decreasing N/O ratios, up to $10^4$ times lower compared to that of Titan's. Even for low N/O ratios, our models are always \ce{N2} dominated.

\subsubsection{Planetary Properties \& Thermal Profiles}
For our models, we take 55 Cnc e as an example of a USP super-Earth. We adopt a planetary mass of $8.59 \pm 0.4$ $M_{\earth}$ and a radius of $1.947 \pm 0.038$ $R_{\earth}$, derived by \citet{Crida_2018}. We construct two general thermal profiles representing the maximum ($T= 2709 \substack{+129 \\ -159}$ K) and minimum ($T= 1613 \substack{+118 \\ -131}$ K) hemisphere-averaged temperatures derived from phase curve measurements \citep{Angelo_2017}. These authors have shown that 1.4 bars of photospheric pressure is needed to explain the derived heat redistribution efficiency value. The profiles are constructed analytically, following a dry-adiabatic profile from the surface of 1.4 bars to 0.1 bars and continuing in isothermal fashion until $10^{-7}$ bar \citep{Morley_2017,Miguel_2019}. This is done using 120 equally spaced log atmospheric layers. The respective temperature profiles are embedded in Figures \ref{fig:chemistry_abundances_2709} and \ref{fig:chemistry_abundances_1613}.

These temperature values represent the possible extremes of the 55 Cnc e and not necessarily the true observables. While this is rather a crude approximation, it is sufficient for the exploration of possible atmospheric compositions in this paper. For a more realistic scenario, the temperature profiles should be constrained using self-consistent radiative transfer calculations, where the profiles are iterated from the given molecular abundances until radiative equilibrium is achieved, which is out of scope of this work.

\subsubsection{Eddy and Molecular Diffusion}
The intermediate, optically thick regions of an atmosphere, may not necessarily follow the thermal equilibrium approximation. If, in an atmospheric region, the dynamical timescale is lower than the chemical timescale, it will become well-mixed. For such super-Earths, with atmospheres of $\sim$1 bar, the well-mixed regions can coincide with the pressure regions that are probed during transmission and emission spectroscopy \citep{Hu_2014}. Thus constraining the parameters of vertical transport is crucial for correctly interpreting observations.

We simulate vertical transport of species by using two different physical processes, eddy diffusion and molecular diffusion. Eddy diffusion is more dominant at intermediate pressure levels, molecular diffusion is more relevant at higher atmospheric altitudes, or lower pressures.

Eddy diffusion is represented by the free parameter K\textsubscript{zz}. Its value can be estimated using scaling relationships from the free-convection and mixing-length theories \citep{Gierasch_1985,Visscher_2010,Hu_2014,Morley_2015}. A crude, commonly used, estimation is taken as the $v_{rms}l$ relationship, where $v$ is the root-mean-square vertical wind velocity obtained from GCMs and $l$ is mixing length, which is taken to be equivalent to the vertical scale height of the atmosphere, but can be a fraction of it \citep{SMITH_1998,Lewis_2010,Moses_2011}. Assuming $l$ of  1 - 10\textsuperscript{4} cm \citep{de_pater_2010},  this results in K\textsubscript{zz} between 10\textsuperscript{6} and 10\textsuperscript{10} cm\textsuperscript{2} s\textsuperscript{-1} for terrestrial planets. However, this may be a crude approximation as the root-mean-square method can result in an overestimation of the K\textsubscript{zz} coefficient  \citep{Parmentier_2013}. For gas giants, K\textsubscript{zz} can also be quantified using theoretical GCM modelling \citep{Parmentier_2013,Charnay_2015,Madhusudhan_2016} or observations \citep{de_pater_2010}. For super-Earths, the lack of observations makes such estimations, for now, unfeasible. Therefore, we adopt a constant K\textsubscript{zz} value of 10\textsuperscript{8} cm\textsuperscript{2} s\textsuperscript{-1} for most of our presented model atmospheres and additionally explore the effects of K\textsubscript{zz} values ranging from 10\textsuperscript{6} to 10\textsuperscript{10} cm\textsuperscript{2} s\textsuperscript{-1} for elemental compositions with C/O ratios between 2.8 and 1.0.

% Paragraph about molecular diffusion
For a planet like 55 Cnc e, molecular diffusion is not expected to have a significant impact on the structure of the visible part of the atmosphere \citep{de_pater_2010,Hu_2014}. This is because the homopause, a region where eddy diffusion and molecular diffusion coefficients are equal for a specific species, is at a pressure level several orders of magnitude lower than the optically thick layers. 

We use the molecular diffusion coefficients D\textsubscript{zz} for each species that are constructed for \ce{N2} dominated atmospheres using the relation

\begin{equation}
D_{zz}=\frac{7.34 \times 10^{16} T^{0.75}}{n} \sqrt{\frac{16.04}{m_{i}}\left(\frac{m_{i}+28.014}{44.054}\right)}
\label{eq:diffusion}
\end{equation}
where $T$ is the atmospheric temperature at a particular pressure level, $n$ is the total number density and $m_{i}$ is the mass of the diffusing species \citep{Banks_1973,Moses_2000}. The expression is derived by taking the \ce{CH4}-\ce{N2} binary diffusion coefficient and scaling it to X-\ce{N2} mixtures with reduced masses of X and \ce{N2} \citep[see][Ch. 15.3]{Banks_1973}.

\subsection{Chemical Models}

UV photolysis, atmospheric escape, condensation, molecular diffusion, mixing, sedimentation, surface emission, can all play a significant role in controlling the chemical composition of the atmosphere. Unlike the atmospheres of hot Jupiters, where the thermochemical equilibrium is obtained at pressures higher than several bar \citep{Moses_2011,Miguel_2014}, the low pressure atmospheres of hot super-Earths are not expected to reach such a state. In this paper we use chemical kinetics to study the temporal evolution of super-Earth atmospheres. We focus on exploring impact of photodissociation, molecular and eddy diffusion. We note that other effects, such as condensation and surface exchange, as well as atmospheric escape could impact the chemistry and will be considered in the follow up studies (see Section \ref{sec:discussion}).

To generate these model atmospheres, we use the open source chemical kinetics code VULCAN\footnote{https://github.com/exoclime/VULCAN} \citep{Tsai_2017} which solves the one-dimensional continuity equation for 4 different elements (\ce{N}, \ce{H}, \ce{C}, \ce{O}) and over 50 molecular species. %(\ce{H2O}, \ce{OH}, \ce{H2}, \ce{CH}, \ce{CH2}, \ce{CH3}, \ce{CH4}, \ce{C2}, \ce{C2H2}, \ce{C2H}, \ce{C2H3}, \ce{C2H4}, \ce{C2H5}, \ce{C2H6}, \ce{CO}, \ce{CO2}, \ce{CH2OH}, \ce{H2CO}, \ce{HCO}, \ce{CH3O}, \ce{CH3OH}, \ce{CH3CO}, \ce{O2}, \ce{H2CCO}, \ce{HCCO}, \ce{NH}, \ce{CN}, \ce{HCN}, \ce{NO}, \ce{NH2}, \ce{N2}, \ce{NH3}, \ce{N2H2}, \ce{N2H}, \ce{N2H3}, \ce{N2H4}, \ce{HNO}, \ce{H2CN}, \ce{HNCO}, \ce{NO2}, \ce{N2O}, \ce{C4H2}, \ce{CH3NH}, \ce{CH2NH}, \ce{CH2NH2}, \ce{CH3NH2}, \ce{CH3CHO}, \ce{HNO2}, \ce{O1D}, \ce{bCH2}).
We employ the reduced NCHO network from VULCAN, consisting of 617 reactions (including reversed), with an additional 35 photodissociation reactions. The forward  reaction rates are obtained mostly from the NIST database\footnote{http://kinetics.nist.gov/kinetics/}. The reverse reaction rates are computed using NASA polynomial thermodynamics data\footnote{http://garfield.chem.elte.hu/Burcat/burcat.html}\citep[see][for the full, non-photochemical reactions list and a more detailed description of the model]{Tsai_2017}.

VULCAN is coupled with the equilibrium chemistry code FastChem\footnote{https://github.com/exoclime/FastChem} \citep{Stock_2018}, which we use to calculate the initial molecular abundances in our atmospheres. FastChem takes a semi-analytical approach of solving the law of mass action and element conservation equations for the composition of a gas phase system \citep[see][]{Stock_2018}. The coupled FastChem version is modified to match the NASA polynomial thermodynamics data and the species list present in VULCAN.

\subsubsection{Stellar Spectrum \& Photochemistry}
\label{spectra}
For our photochemical calculations we have constructed a synthetic spectrum for 55 Cancri, a 0.98 solar radii star with T\textsubscript{eff}$\sim$5200 K \citep{Bourrier_2018,Yee_2017,von_Braun_2011}. It is composed of the synthetic ATLAS spectrum for a star with T$_{eff}$ = 5250 K, with coadded UV spectra extensions from the International Ultraviolet Explorer observations of a similar star - HD 10780 \citep{Kurucz_1979,Rugheimer_2013}. It is further combined with solar chromospheric far-UV emission up to 115.5 nm from the VPL\footnote{http://depts.washington.edu/naivpl/content/spectral-databases-and-tools} database. We use a fixed zenith star angle of 48 degrees as in \citet{Moses_2011} and a uniform wavelength binning of 0.1 nm. 

The photodissociation rates are calculated using
\begin{equation}
    k(z)=\int q(\lambda) \sigma_{d}(\lambda) J(\lambda, z) d \lambda
    \label{eq:photodissociation}
\end{equation}
where, $q(\lambda)$ is the quantum yield, defined as the rate of occurrence of certain photodissociation products per absorbed photon, $\sigma_{d}(\lambda)$ is the absorption cross-section, $J(\lambda, z)$ is the calculated actinic flux from both, direct stellar beam and scattered radiation contributions, with a two-stream approximation from \citet{Malik_2017}.

The photodissociation pathways and their quantum yield values can be found inside the NCHO photo network file which is part of the latest VULCAN version (Tsai et al. (in prep)). Due to its importance, we additionally add CN photodissociation to this network. For the cross sections, we use the Leiden\footnote{https://home.strw.leidenuniv.nl/~ewine/photo}, PHID RATES\footnote{https://phidrates.space.swri.edu} and MPI-Mains UV/VIS Spectral Atlas\footnote{http://satellite.mpic.de/spectral\_atlas} values.

\subsection{Synthetic Spectra}
\subsubsection{Radiative Transfer code petitRADTRANS}
\label{petitradtrans}
\begin{figure}
	\includegraphics[width=\columnwidth]{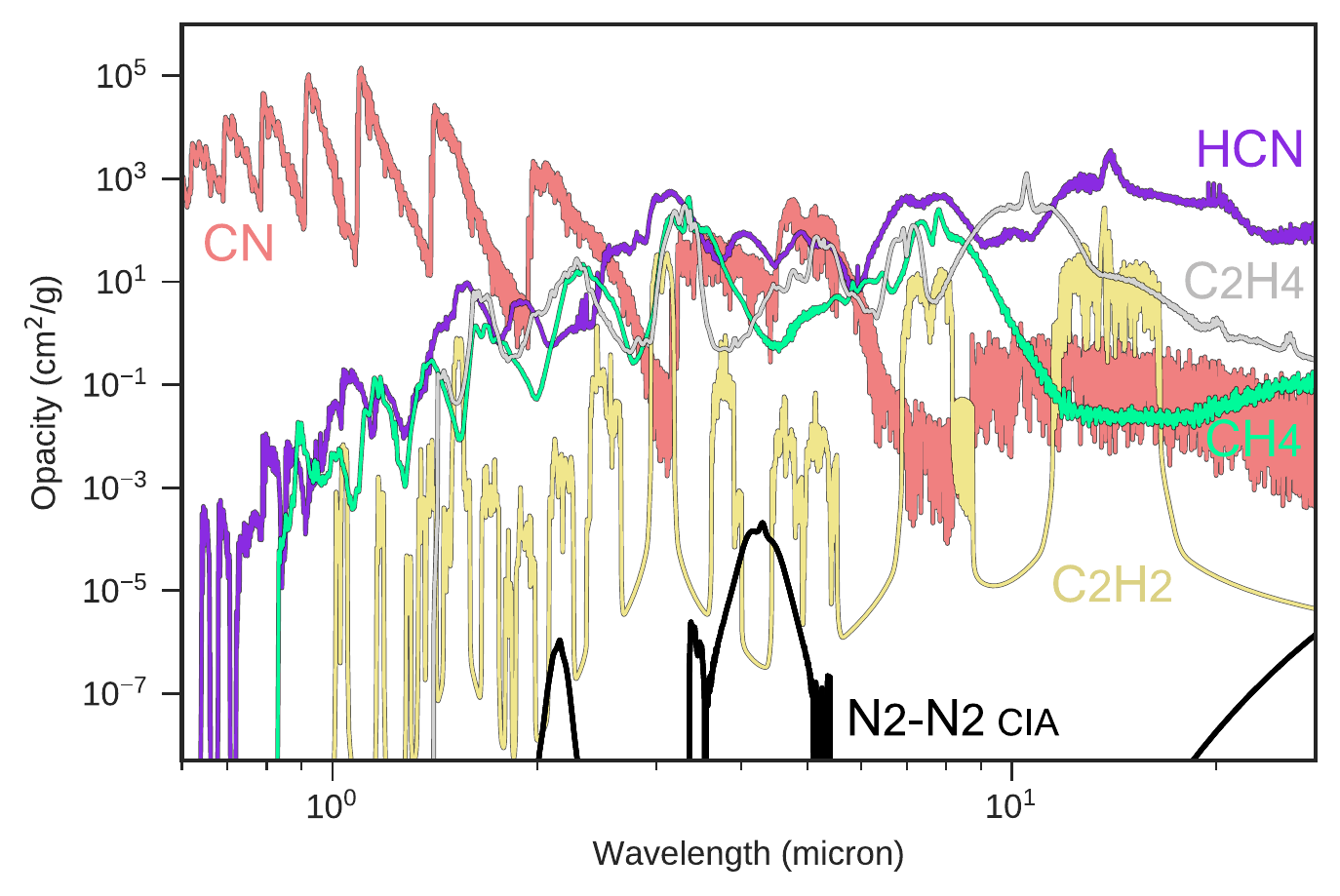}
    \caption{Line opacities of species more important for C/O > 1.0 compositions in our models. Modelled at a temperature of 2709 K and pressure of 1.4 bar. Included is the low temperature opacity data of \ce{N2-N2} collision-induced absorption from the HITRAN database. For each species, the opacity is calculated assuming 100\% abundance and shown at a resolution of $\lambda/\Delta\lambda = 1000$. Different colours indicate different species.}
    \label{fig:opacities}
\end{figure}

\begin{figure}
	\includegraphics[width=\columnwidth]{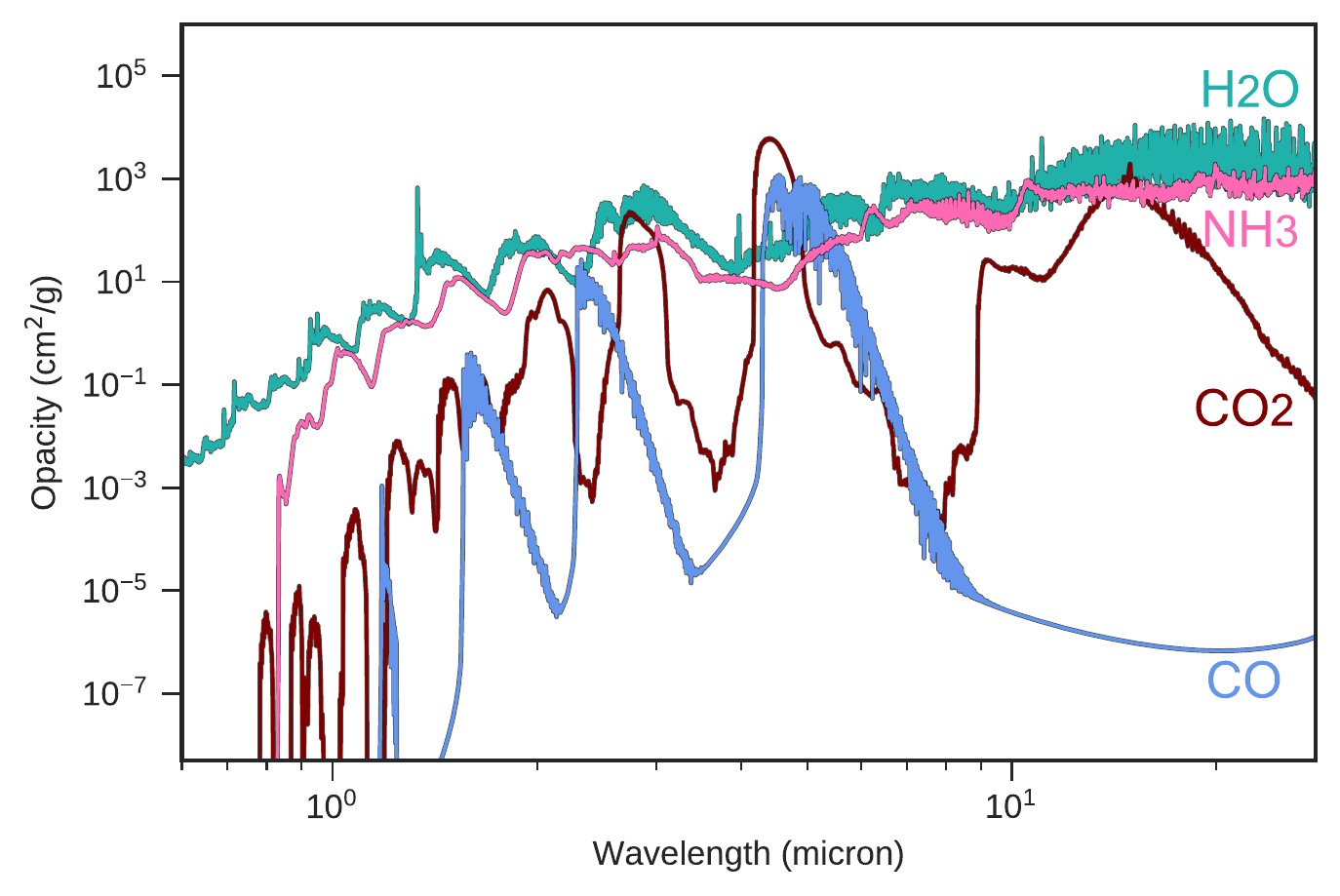}
    \caption{Line opacities of species that are more important for C/O < 1.0 compositions in our models. Shown for a temperature of 2709 K and pressure of 1.4 bar. Included is the low temperature opacity data of \ce{N2-N2} collision-induced absorption from the HITRAN database. For each species, the opacity is calculated assuming 100\% abundance and shown at a resolution of $\lambda/\Delta\lambda = 1000$. Different colours indicate different species.}
    \label{fig:opacities_2}
\end{figure}

Our synthetic emission and transmission spectra are computed using the open source radiative transfer code petitRADTRANS\footnote{http://gitlab.com/mauricemolli/petitRADTRANS} \citep{Molliere_2019}. We use the low resolution mode of $\lambda/\Delta\lambda = 1000$, which follows the correlated-k approximation. We take as line opacities: \ce{H2}, \ce{H2O}, \ce{HCN}, \ce{C2H2}, \ce{C2H4}, \ce{CO}, \ce{CO2},  \ce{CH4}, \ce{CN}, \ce{NH3}, \ce{OH}, \ce{CH}, \ce{NH} and \ce{NO}. Respective sources of the opacities can be found in the Table 2 of \citet{Molliere_2019}, and the code documentation page\footnote{https://petitradtrans.readthedocs.io}, where petitRADTRANS is described at length. The reference list for added opacities, which were not originally part of petitRADTRANS, of \ce{C2H4}, \ce{CH}, \ce{NO}, \ce{NH} and \ce{CN} can be found in Table \ref{tab:opacity_sources}. Added opacities for \ce{CN} and \ce{NH} are calculated in this work using the ExoCross code with the line lists and partition functions referenced in the table. For the transmission spectra, we include Rayleigh scattering cross-sections of \ce{N2}, \ce{H2} and \ce{He}. We also include collision induced absorption (CIA) of \ce{N2}-\ce{N2}, \ce{N2}-\ce{H2}, \ce{H2}-\ce{H2} and  \ce{H2}-\ce{He} pairs. CIA opacities for \ce{N2}-\ce{N2} and \ce{N2}-\ce{H2} pairs were added from the HITRAN database \citep{Borysow_1987,Gruszka_1997,Samuelson_1997,Gustafsson_2001,Drouin_2017}. In Figures \ref{fig:opacities} and \ref{fig:opacities_2} we show line opacities for a selection of species at a temperature of 2709 K and a pressure of 1.4 bar. These opacities are shown assuming 100\% abundance for each individual species, which do not represent abundances expected from chemistry. We note that the existing \ce{N2}-\ce{N2} and \ce{N2}-\ce{H2} data has been only compiled for temperatures up to 400 K, the implications of this are further discussed in Section \ref{sec:discussion}. 

\begin{table}
 \caption{Sources of selected opacities}
 \label{tab:opacity_sources}
 \begin{tabular}{llll}
  \hline
  Species & Source & Isotope & Line list\\
  \hline
  \ce{CN} & Own$^*$ & 12C-14N & Yueqi$^1$\\
  \ce{CH} & Opacity World$^{**}$ & 12C2-1H & Bernath$^2$\\
  \ce{C2H4} & Opacity World & 12C2-1H4 & MaYTY$^3$\\
  \ce{NH} & Own$^*$ & 14N-1H & Bernath$^4$\\
  \ce{NO} & Opacity World & 14N-16O & NOname$^5$\\
  \hline
    \multicolumn{4}{p{7.8cm}}{\footnotesize$^*$ Calculated using ExoCross code \citet{Yurchenko_2018}. Partition functions from ExoMol \citep{tennyson_2016}.}\\
    \multicolumn{4}{p{7.8cm}}{\footnotesize$^{**}$\citet{Grimm_2015}, http://opacity.world}\\
    \multicolumn{4}{p{7.8cm}}{\footnotesize$^1$\citet{Brooke_2014}, $^2$\citet{Masseron_2014}, $^3$\citet{Barry_2018}, $^4$\citet{Brooke_2014b,Brooke_2015,Fernando_2018}, $^5$\citet{Wong_2017}}\\
 \end{tabular}
\end{table}

As petitRADTRANS input, we relay all of the atmospheric parameters and the molecular abundances from the output of VULCAN. In order to calculate star-to-planet ratio for the emission spectra, we use a PHOENIX generated spectrum for the temperature of 5196 K. We additionally explore the star-to-planet ratio deviations when using the mostly empirical 55 Cancri spectrum by \citet{Crossfield_2012}.

\subsubsection{JWST instrumentation noise model - PandExo}

We assess the observability of our modelled atmospheres with JWST instrumentation using the PandExo package by \citet{Batalha_2017}, which is built around the core of the Space Telescope Science Institute's Exposure Time Calculator - Pandeia\footnote{http://jwst.etc.stsci.edu}. We simulate two NIRCam's grism (F322W2 and F444W) and the MIRI LRS observing modes in both, emission and transmission, using 4 eclipses or 4 transits, respectively. Combined, these instruments cover the observing range from 2.4 to 12 $\mu$m. Out of all the JWST observing modes, NIRCam's F444W and MIRI LRS are least affected by saturation for bright targets, such as 55 Cancri (J = 4.59). NIRCam's F322W2 observing mode is a lot more sensitive to saturation and would most likely prove to be ineffective for very bright targets. In order to avoid saturation as much as possible, NIRCam's observing modes are taken with the lowest subarray modes - SUBGRISM64 of 64x2048 pixels which have the maximum possible readout speed. For MIRI LRS we use the slitless prism mode, which has several orders of magnitude brighter saturation limit than the slit mode. We adopt systematic noise floor values estimated by \citet{Greene_2016}, 30 and 50  ppm for NIRCam and MIRI LRS, respectively. The real values, just like saturation limits, will not be known until after launch of JWST and could very well improve, especially if observational techniques such as spatial scanning are eventually employed \citep{Tsiaras_2016b}.

\label{sec:pandexo}

\section{Results}
\label{sec:results}
\subsection{Chemistry}

In this section we present the results of the photochemical kinetics code. We show a selection of highly-absorbing species as a function of C/O, N/O or atmospheric pressure. In addition, we demonstrate the effects of photodissociation and varying strengths of mixing on the molecular composition of our modelled atmospheres.

\subsubsection{Chemical Composition}
\begin{figure*}
	\includegraphics[width=0.95\textwidth]{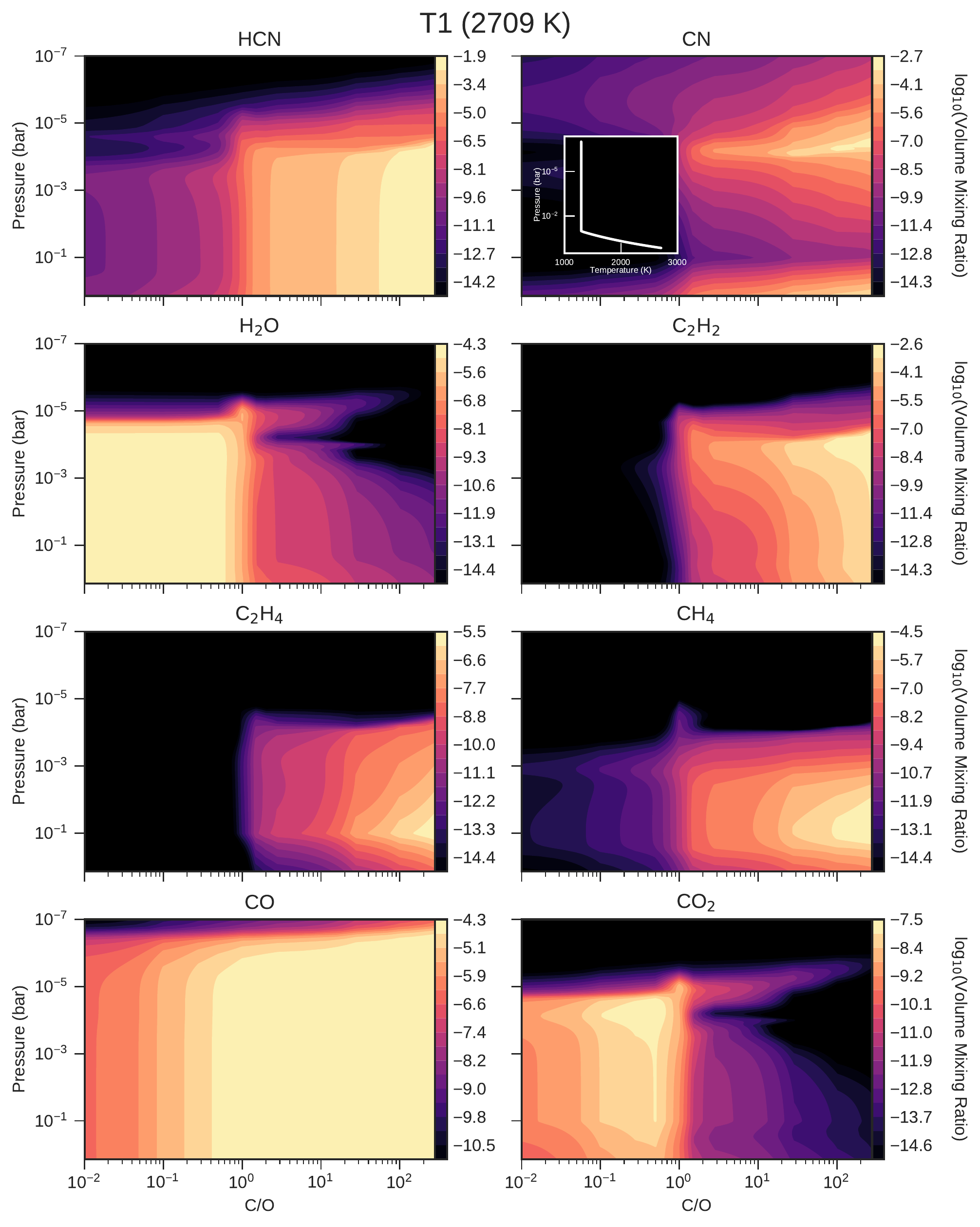}
    \caption{Volume mixing ratios of major chemical species with important spectral features in a nitrogen dominated atmosphere of a super-Earth resembling 55 Cnc e. Abundances are shown as a function of the C/O ratio and atmospheric pressure. The VMR are calculated using chemical kinetics with a temperature profile indicated in the embedded plot (upper right), which represents the maximum hemisphere-averaged temperature (T1) from the phase curve measurements for 55 Cnc e \citep{Angelo_2017}. All cases shown with K\textsubscript{zz} = $10^{8}$ cm\textsuperscript{2} s\textsuperscript{-1} and including photochemistry. The initial elemental composition is taken to be that of Titan's atmosphere. The background molecular atmosphere in all cases is composed of \ce{N2} and \ce{H2}.}
    \label{fig:chemistry_abundances_2709}
\end{figure*}

\begin{figure*}
	\includegraphics[width=0.95\textwidth]{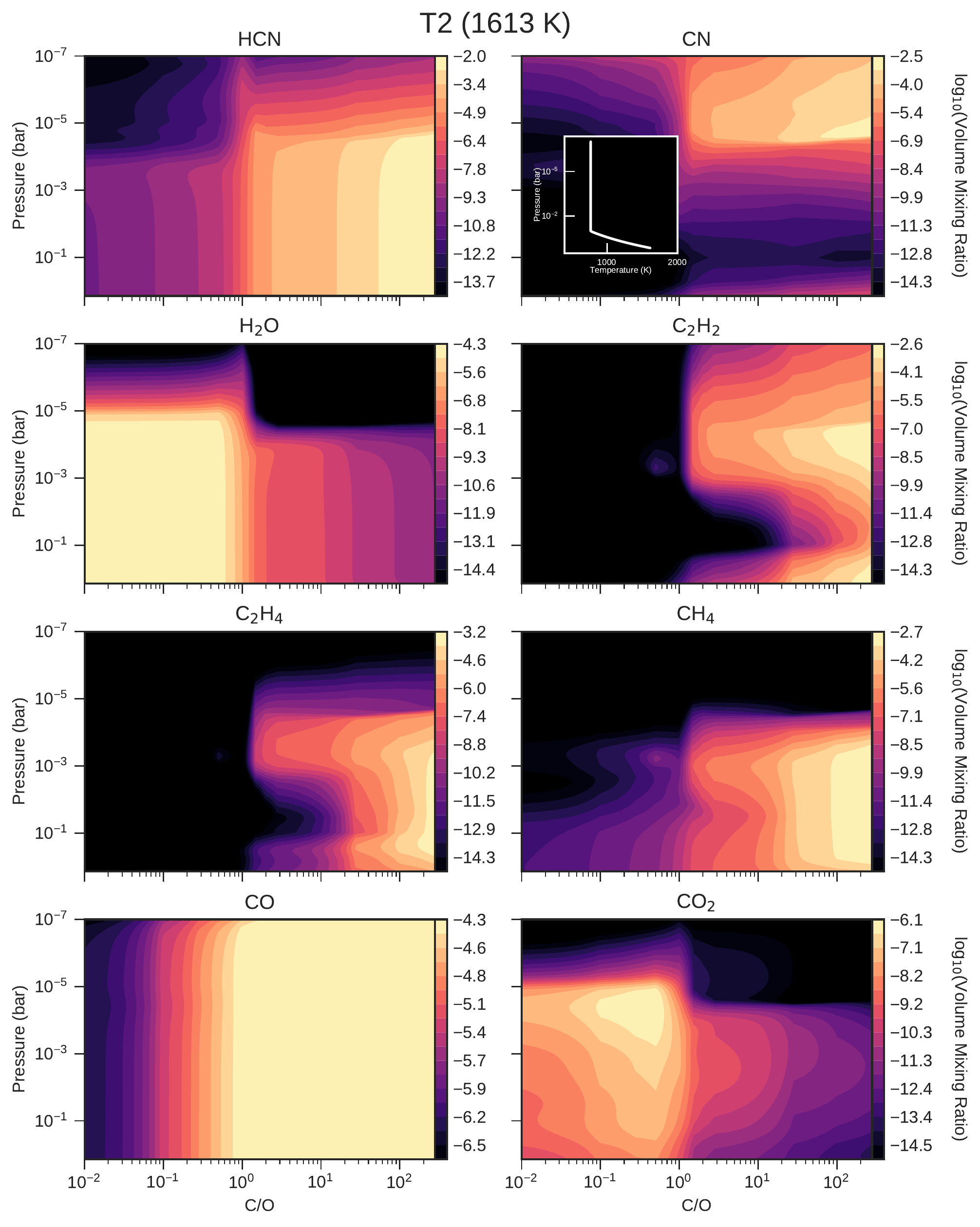}
    \caption{Abundances of chemical species with important spectral features in hot super-Earth atmospheres with nitrogen as the main component. VMR are shown as a function of the C/O ratio and atmospheric pressure. The thermal profile used for chemical kinetics calculations is shown in the embedded plot (upper right) and represents the minimum hemisphere-averaged temperature (T2) from the phase curve measurements of 55 Cnc e \citep{Angelo_2017}. All cases shown with K\textsubscript{zz} = $10^{8}$ cm\textsuperscript{2} s\textsuperscript{-1} and including photochemistry. The initial elemental composition is taken to be that of Titan's atmosphere. The background molecular atmosphere in all cases is composed of \ce{N2} and \ce{H2}.}
    \label{fig:chemistry_abundances_1613}
\end{figure*}

% CO > 1.0 paragraph
Figures \ref{fig:chemistry_abundances_2709} and \ref{fig:chemistry_abundances_1613} show the volume mixing ratios (VMR), as a function of C/O and pressure, of species relevant for spectroscopy observations. Each figure depicts the results for a different thermal profile, which is indicated by the embedded plot. From here on we refer to two temperature profiles, that is, the maximum hemisphere-averaged and minimum hemisphere-averaged, as T1 and T2, respectively. 

\paragraph{Compositions with varied C/O:} The modelled compositions in this section are with Titan's N/O of $3.85 \times 10^{4}$. Regardless of the temperature range, we see that the NCHO molecular composition of an atmosphere heavily depends on its carbon-to-oxygen ratio. In particular, almost all of the chemical species that contain carbon or oxygen, experience a sharp abundance change at C/O $\sim$ 1.0, similar to what happens in hydrogen dominated atmospheres \citep{Madhusudhan_2012,Moses_2013a,Venot_2015,Molliere_2015,Rocchetto_2016}. The notable exception is \ce{CO}, which is consistently abundant, only varying few orders of magnitude for the C/O ratios considered. This is because in nitrogen based, carbon-rich atmospheres, \ce{CO} is not only the major carbon carrier, but also a major oxygen carrier. At C/O ratios above unity, its abundance is only limited by the amount of oxygen available in the atmosphere. Once the C/O ratio drops below unity, the abundance of \ce{CO} starts to slowly reduce, becoming suppressed by the availability of carbon.

An increase in the C/O ratio above 1.0 results in an increased abundances of all carbon bearing species. If nitrogen is introduced to the system, an atmosphere containing even a small amount of hydrogen ($\sim 10^{-6}$ ) will tend to form hydrogen cyanide (HCN). Due to large quantity of \ce{N} and sufficient \ce{H} in nitrogen dominated atmospheres, \ce{HCN} can quickly become the major carbon carrying species when oxygen abundances are low, which is true above C/O ratios of 2.0. \ce{HCN} is a very strong absorber, and even in small quantities (VMR > $10^{-6}$), it can dominate the spectral fingerprint of a super-Earth's atmosphere (see Section \ref{sec:spectra}). 

Because of the lack of hydrogen (our base model has an H mass fraction of 0.03, see Section \ref{sec:meth_atmos_comp}), other carbon carriers, like \ce{CH4} and hydrocarbons such as \ce{C2H2}, \ce{C2H4}, \ce{C2H6}, have VMR above $\sim10^{-8}$ only at C/O ratios higher than 2.0, and quickly decrease in VMR as the ratio is reduced.  Even though \ce{C2H2} can become quite abundant, it never dominates over \ce{HCN}. Only in non-nitrogen, or high hydrogen-content atmospheres would it become a major species. \ce{C2H4} becomes a significant species only in cooler temperatures (T2). Methane is also more stable in lower temperatures, though is still relatively low in abundance for both, T1 and T2.  \ce{CH4} is generally more than two orders of magnitude lower to that of \ce{HCN} or \ce{CO} for C/O < 3.0. We find that in many cases \ce{CH4} becomes significant only in thermal equilibrium, but largely decreases in abundance when kinetics are applied. 

For atmospheres with C/O < 1.0, the compositions are largely dominated by \ce{N2} and \ce{H2}, with \ce{H2O}, \ce{CO}, \ce{CO2} as strong absorbers. \ce{HCN} and hydrocarbons drop below the detectable levels. Water and carbon dioxide tend to show similar increases or decreases as they are both chemically favoured and interlinked with each other in oxygen-rich atmospheres \citep{Moses_2011,Venot_2015}. \ce{H2O}, \ce{CO2} and \ce{CO} are not strongly affected by the temperature difference between the two profiles. Although \ce{CO2} is found to have slightly increased VMR for lower temperatures.

\begin{figure}
	\includegraphics[width=\columnwidth]{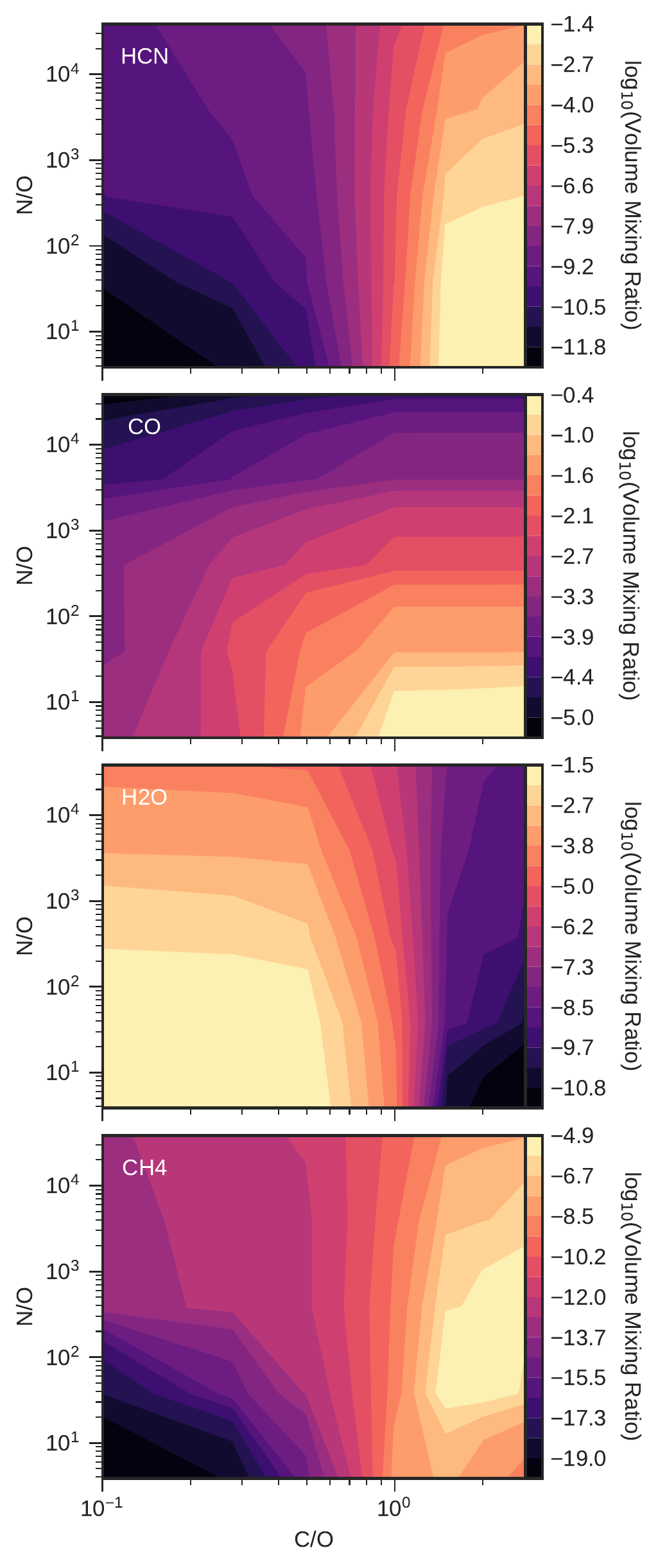}
    \caption{The change in VMR as a function of C/O and N/O in N-dominated atmospheres similar to 55 Cnc e. Abundances are calculated using chemical kinetics and represented as averaged values of the photospheric region between 1.4 - $10^{-4}$ bar, for the maximum hemisphere-averaged temperature profile (T1) (embedded in Fig. \ref{fig:chemistry_abundances_2709}).}
    \label{fig:chemistry_stack_2709}
\end{figure}

\paragraph{Compositions with varied N/O:} So far we have shown compositions with Titan's N/O of $3.85 \times 10^{4}$. We have also explored the changes that occur when the N/O ratio is varied. Figure \ref{fig:chemistry_stack_2709} shows how \ce{HCN}, \ce{CO}, \ce{H2O} and \ce{CH4} abundances change with decreasing N/O ratio, while still keeping nitrogen the major component of the atmosphere. The shown values are averages of the photospheric region, which we find to be, in most cases, from $\sim$1.4 to $10^{-4}$ bar. Due to photochemistry, certain species can accumulate enough in the upper regions of the atmosphere ($\lesssim10^{-4}$ bar) to become strong absorbers. \ce{CN}, a major nitrogen carrier that is easily produced by photochemistry, is an example of this in models with C/O ratios above unity. In Fig. \ref{fig:chemistry_stack_2709} we only show results for T1, as T2 has similar behaviour. 

Our atmospheric models have a fixed hydrogen abundance. To maintain the total mass fraction of 1 when reducing the abundance of nitrogen (vary N/O), we have to increase carbon or oxygen to compensate for this. Because we maintain the C/O ratio when we vary N/O, both C and O are always adjusted with the same proportion. Thus the general effect is that formation of most species that need C or O will be enhanced if N/O is reduced. If the N/O ratio is reduced far enough ($\lesssim10^2$), the significantly increased carbon budget will allow for extremely efficient formation of \ce{C4H2}, which quickly becomes one of the major constituents in the atmosphere. Because our used network truncates at hydrocarbon species containing two carbon atoms, the abundance of \ce{C4H2} could be overestimated. Using a larger chemical scheme for high C/O ratio compositions, such as the one of \citet{Venot_2015} would result in a more accurate estimation. Large quantities of \ce{C4H2} would likely polymerise to form polycyclic aromatic hydrocarbons (PAHs) and hazes \citep{Zahnle_2016}, flattening the observed spectrum. Most of the added oxygen will go into production of \ce{CO}, in N/O $\lesssim10^2$ cases pushing it close to that of \ce{N2} abundances.  The increased O/H ratio will deplete species like \ce{CH4} to a point where it is no longer present in the spectrun, even at C/O ratios $\gg$ 1.0.

Figure \ref{fig:chemistry_stack_2709} shows that \ce{HCN} abundance, at C/O ratios larger than $\sim$1.3, is heavily dictated by the supply of nitrogen. When the C/O ratio starts to reduce below this value, \ce{HCN} abundance becomes more limited by the availability of carbon, which it competes for with other major species, e.g. \ce{CO}. Hydrocarbons and \ce{CN} will tend to follow the behaviour of \ce{HCN}. As C/O is reduced further, the atmosphere becomes filled with oxidised species, e.g, \ce{CO2} and \ce{H2O}. Both of these species heavily rely on the availability of oxygen, which grows with N/O reduction. When the N/O ratio is high, \ce{CO}, being the major carbon and oxygen carrier, is not strongly influenced by the C/O ratio itself, but rather by the combined amount of both, carbon and oxygen. Decreasing N/O will cause \ce{CO} to become more dependant on the C/O ratio as it decreases below unity.

\subsubsection{Effect of photochemistry and vertical transport}

\begin{figure*}
	\includegraphics[width=\textwidth]{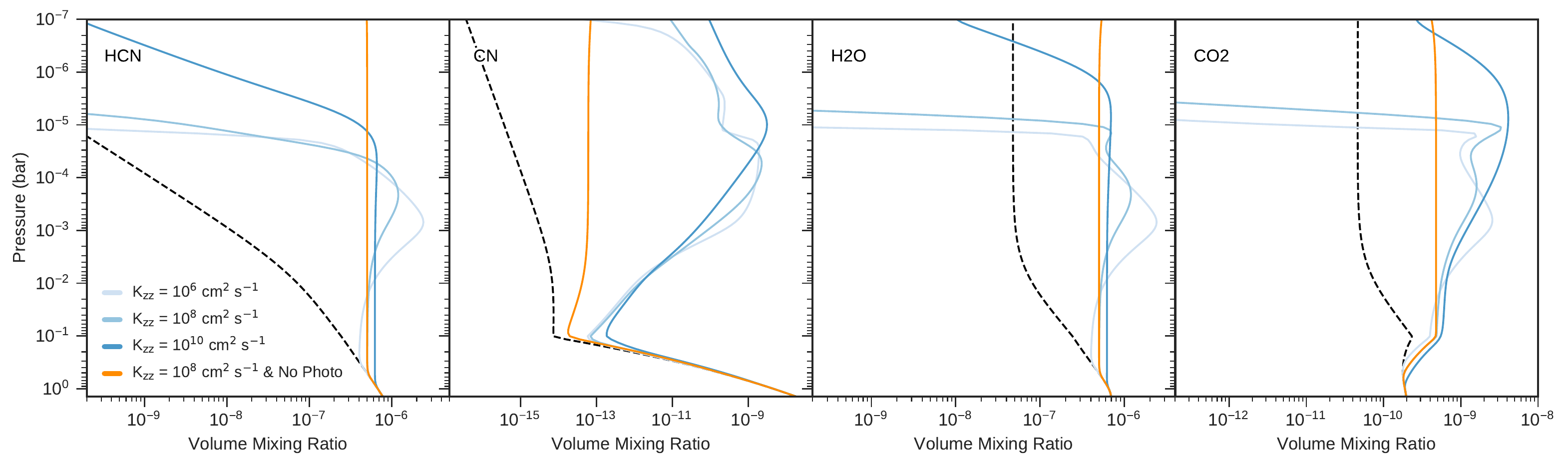}
    \caption{Volume mixing ratios of selected species as a function of pressure (C/O = 1.0, using T2 for 55 Cnc e).  The dashed curves represent chemical equilibrium values. Orange curves show the effects of vertical mixing and molecular diffusion without active photochemistry. Differently shaded blue curves show how atmospheric abundances change with increasing eddy diffusion coefficient.The temperature profile is embedded in Fig. \ref{fig:chemistry_abundances_2709}.}
    \label{fig:kzz_2709}
\end{figure*}

\paragraph{Photochemistry:} As illustrated in Figure \ref{fig:kzz_2709}, the atmosphere is heavily perturbed by the effects of photochemistry and vertical mixing. In upper atmospheric regions, where photochemistry is significant, that is approximately above $10^{-2}$ bar, atmospheric abundances are shifted away from equilibrium values. 

In the highest regions (P $\lesssim10^{-5}$ bar), depending on the vertical mixing coefficient, various species will be strongly dissociated. This region will become dominated by atomic species, reducing its mean molecular weight. Many of the species containing hydrogen will be photolysed, potentially allowing atomic hydrogen to escape the atmosphere over the lifetime of the planet (see discussion in Section \ref{sec:discussion_hydrogen_escape}). Whether dissociation would allow heavier species, e.g. \ce{N} or \ce{C} to escape, is largely unknown and not explored in this paper.

For the region between $10^{-2}$ and $10^{-5}$ bar, the photolysis of \ce{CO}, amongst other species, will produce high abundances of available \ce{C} and \ce{O} atoms. Because thermal decomposition of \ce{CN} is extremely slow at low temperatures and low pressures, most of the produced carbon atoms will go into \ce{CN}. In fact, due to the stability of \ce{CN} molecules, its abundance will increase several orders of magnitude when photochemistry is turned on. \ce{CN} VMR can even rival that of \ce{HCN} at C/O ratios much larger than unity. As shown in Fig. \ref{fig:kzz_2709}, \ce{HCN}, along with many other species will show increased abundances at around $\sim10^{-3}$ bar in a form of a ``bump", which is dependant on the strength of K\textsubscript{zz}. For \ce{HCN}, this is initiated by the photolysis of \ce{CO} through the following path:
\begin{description}
\item 2[\ce{CO + h$\nu$ -> C + O}]

\item \ce{N2 + h$\nu$ -> N + N}

\item 2[\ce{C + N + M -> CN + M}]

\item 2[\ce{CN + H + M -> HCN + M}]

\item 2[\ce{O + H + M -> OH + M}]

\item 2[\ce{OH + H + M -> H2O + M}]
      \vspace{0.2cm}
      \hrule width0.8\columnwidth
      \vspace{0.2cm}     
\item net: \ce{2CO + N2 + 6H -> 2HCN + 2H2O}
\end{description}
Here, the \ce{CO} photodissociation is the rate-limiting step. This process is similar to that in hydrogen dominated atmospheres on hot Jupiters, with the difference being that \ce{N2} is shielded from photodissociation by highly abundant hydrogen species \citep{Moses_2011}.

\paragraph{Vertical transport:} Vertical mixing is an important effect that can affect the composition of the atmosphere. When the mixing timescale is shorter than the chemical timescale of a species, its chemistry will freeze (quench) and the abundance will be modified according to the strength of the mixing. In Figure \ref{fig:kzz_2709}, we show examples of this quenching process. For \ce{HCN}, \ce{H2O} and \ce{CO2} this occurs approximately at 0.6 bars (K\textsubscript{zz} = $10^{8}$ cm\textsuperscript{2} s\textsuperscript{-1}). As seen in Fig. \ref{fig:kzz_2709}, an increase in vertical mixing strength quenches the species at higher pressures. Strong mixing (such as the case with K\textsubscript{zz} = $10^{10}$ cm\textsuperscript{2} s\textsuperscript{-1} in Fig. \ref{fig:kzz_2709}) will mix molecules into the higher, low pressure regions, reducing the effect of photochemistry.

\subsection{Synthetic Spectra}
\label{sec:spectra}
In this section we show synthetic spectra for selected model atmospheres. The spectra are computed with log(g), R\textsubscript{pl} and T\textsubscript{surf} of 55 Cnc e, as stated in Section \ref{petitradtrans}. For both, transmission and emission, these are shown for a variation of C/O and N/O ratios. We also show additional spectra for models having different eddy diffusion coefficients, as in Figure \ref{fig:kzz_2709}.

\subsubsection{Transmission Spectra}
\begin{figure*}
	\includegraphics[width=\textwidth]{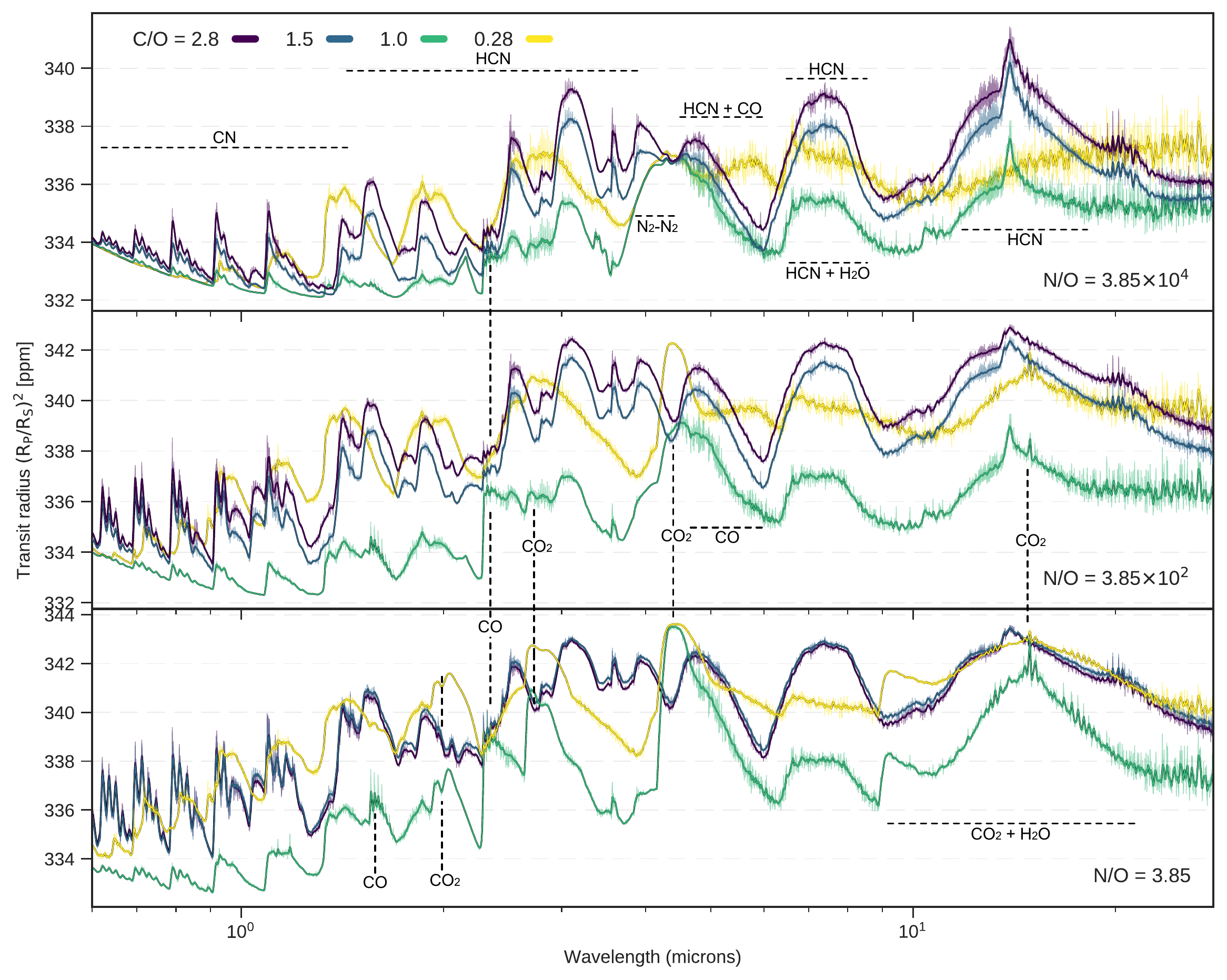}
    \caption{Synthetic transmission spectra corresponding to the maximum hemisphere-averaged temperature profile (T1) of 55 Cnc e. Shown are the disequilibrium results with a K\textsubscript{zz} = $10^{8}$ cm\textsuperscript{2} s\textsuperscript{-1} for range of C/O and N/O ratios. Spectra are highlighted in lower resolution for visibility. Faded spectra are shown at native $\lambda/\Delta\lambda$ = 1000. Each panel represents a different N/O ratio indicated in the lower right corner. Horizontal, dashed lines indicate major contributing species of the broad absorption features. Vertical, dashed lines indicate narrow features. Carbon-rich atmospheres (Purple, blue) are defined in shape by \ce{HCN} and \ce{CN}. C/O = 1.0 (Green) is defined by \ce{H2O}. Low C/O ratio spectra (Yellow) are dominated by the absorption of \ce{H2O}, and by \ce{CO2} for the lowest N/O ratios.}
    \label{fig:transmission_2709}
\end{figure*}

\begin{figure*}
	\includegraphics[width=\textwidth]{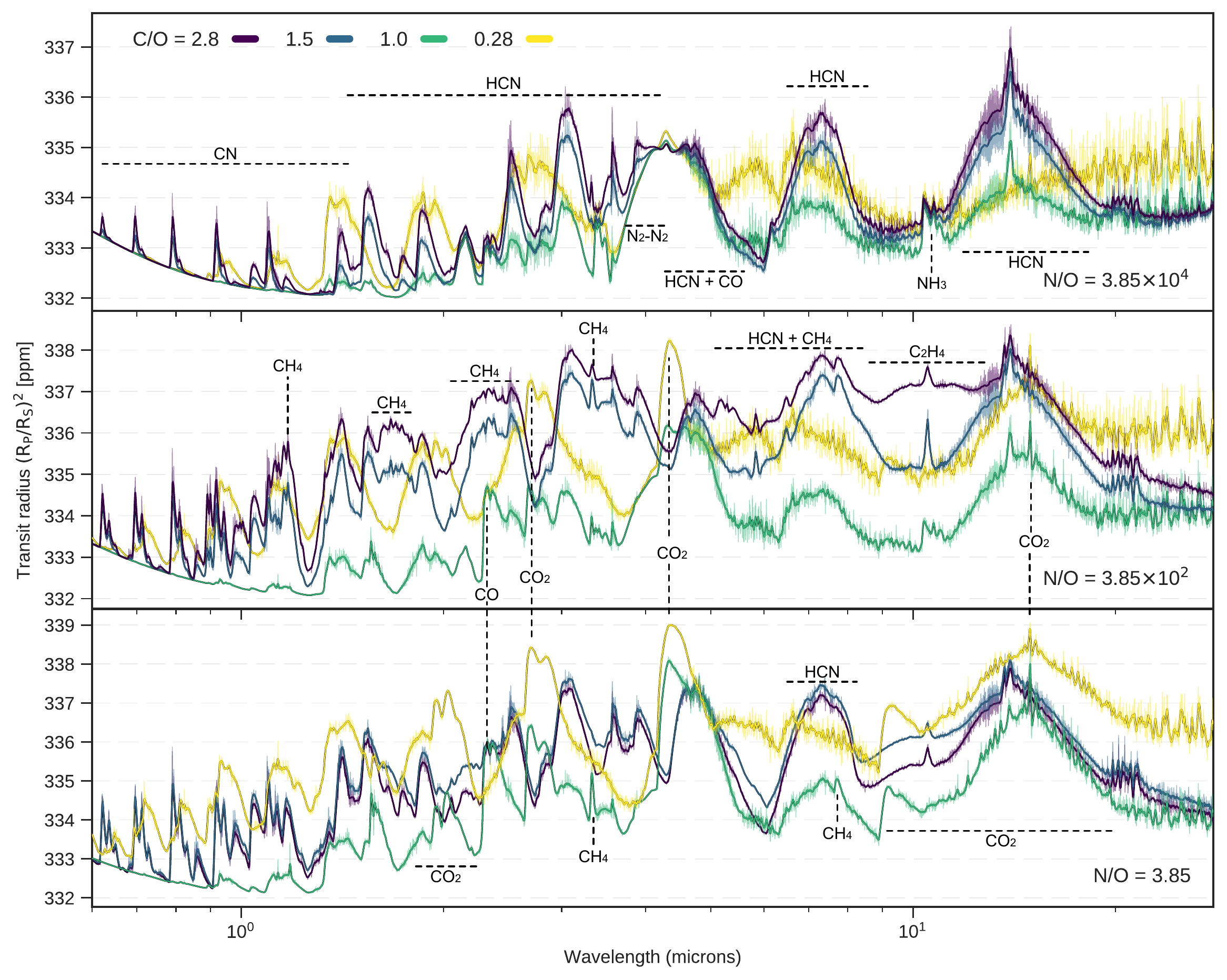}
    \caption{Synthetic transmission spectra depicting disequilibrium results with a K\textsubscript{zz} = $10^{8}$ cm\textsuperscript{2} s\textsuperscript{-1} for range of C/O and N/O ratios. Calculated for the T2 temperature profile.  Horizontal, dashed lines indicate major contributing species of the broad absorption features. Vertical, dashed lines - narrow features. Carbon-rich atmospheres (Purple, blue) are defined in shape by \ce{HCN} and \ce{CN}. C/O = 1.0 (Green) is defined by \ce{H2O}. Low C/O ratio spectra (Yellow) are dominated by the absorption of \ce{H2O}, and by \ce{CO2} for the lowest N/O ratios.}
    \label{fig:transmission_1613}
\end{figure*}

Figures \ref{fig:transmission_2709} and \ref{fig:transmission_1613} depict transmission spectra for T1 and T2 profiles, respectively. We indicate major broad molecular contributors with horizontal, dashed lines. Vertical, dashed lines show narrow features. 

\textit{C/O > 1.0}: With carbon being more abundant than oxygen in the atmosphere, the spectral shape becomes mostly dominated by the presence of \ce{HCN} and \ce{CN}. \textbf{\ce{CN}} shows a series of repeating peaks from 0.6 to 1.5 $\micron$, where it starts to overlap with \ce{HCN}. Both molecules have stronger features with larger C/O ratio. \textbf{\ce{CO}} absorbs at 2.3 and 4.6 $\micron$, with the 4.6 $\micron$ feature being much more significant. At 4.6 $\micron$ \ce{HCN} is also a strong absorber and for high C/O ratios is more dominant than \ce{CO}. Only when the C/O ratio approaches unity does the \ce{CO} become a more dominant absorber than \ce{HCN}. \textbf{\ce{C2H2}} absorption (2.6, 3.1, 7.5 and 13.9 $\micron$) becomes stronger as the N/O ratio is decreased, but because nitrogen is always the main component of the atmosphere, it never overtakes or contributes significantly to the strong presence of \ce{HCN}. We also see slight \ce{N2-N2} CIA absorption at 4.1 $\micron$.

For the lower temperatures (T2) (Fig. \ref{fig:transmission_1613}), differences are several. \ce{CN} features are diminished. \ce{CO} is more dominant at 4.6 $\micron$ than \ce{HCN}. \ce{NH3} has a feature at 10.5 $\micron$, which with decreasing N/O ratio gets overtaken by \ce{C2H4}. The lower temperature range of the T2 profile also allows \textbf{methane} to be abundant enough to appear in the spectra. In Fig. \ref{fig:transmission_1613}, for N/O = $3.85 \times 10^{2}$ and C/O > 1.0, it has multiple features between 1.0 - 3.3 and at 8.0 $\micron$. While the reduction of N/O ratio will provide more available carbon for \ce{CH4} to form, for the lowest N/O ratios, \ce{C4H2} will be a more efficient carbon sink, making methane disappear from the spectra completely. As shown in Fig. \ref{fig:chemistry_stack_2709}, which is also applicable to T2, when C/O > 1.0, \ce{CH4} is most abundant in the mid-ranges of the presented N/O ratios.

\textit{C/O = 1.0}: The main differences from previous cases is that the atmosphere opacity has a minimum at this C/O ratio. Most of the available \ce{C} and \ce{O} go into producing \ce{CO}. The result of this is that other species become heavily diminished, thus producing minimum signal amplitude in the shown wavelength region \citep{Molliere_2015}. The spectrum shape is now mostly defined by \ce{H2O}, with \ce{HCN} features at 3.0, 7.5 and 11.0 - 15.0 $\micron$, and \ce{CO} at 1.6, 2.3, 4.6 $\micron$. At high N/O ratios, \ce{N2-N2} CIA is still present at 4.1 $\micron$. When the N/O ratio is reduced \ce{CO2} becomes increasingly present at 2.0, 2.8, 4.3 and 19.0 - 20.0 $\micron$. At N/O = 3.85, \ce{CO2} becomes the dominant absorber between 9.0 to 20 $\micron$.

For the T2 profile, the changes are similar to the C/O > 1.0 cases with few small differences. Due to the lower carbon abundance, \ce{HCN} is never dominant at 4.6 $\micron$, regardless of our used temperature profile. \ce{C2H4} never appears since \ce{CO2} is now the dominant absorber. Methane is much less prominent, showing only slight peaks at 3.3 and 8.0 $\micron$, even at the lowest N/O ratios as \ce{C4H2} formation is not efficient when C/O = 1.0.

\textit{C/O < 1.0}: Once the ratio drops below unity, the spectra become increasingly dominated by \ce{H2O}. \ce{HCN} features vanish. \ce{CO} is present at 4.6 $\micron$ for higher N/O ratios. As with C/O = 1.0 cases, when the oxygen reservoir increases (N/O reduced), \ce{CO2} becomes increasingly dominant, having features at 2.0, 2.8, 4.3 and 9.0 - 20.0 $\micron$. For very low C/O ratios, the difference between our T1 and T2 profiles are minor as no new species emerge as dominant absorbers.

\begin{figure}
	\includegraphics[width=\columnwidth]{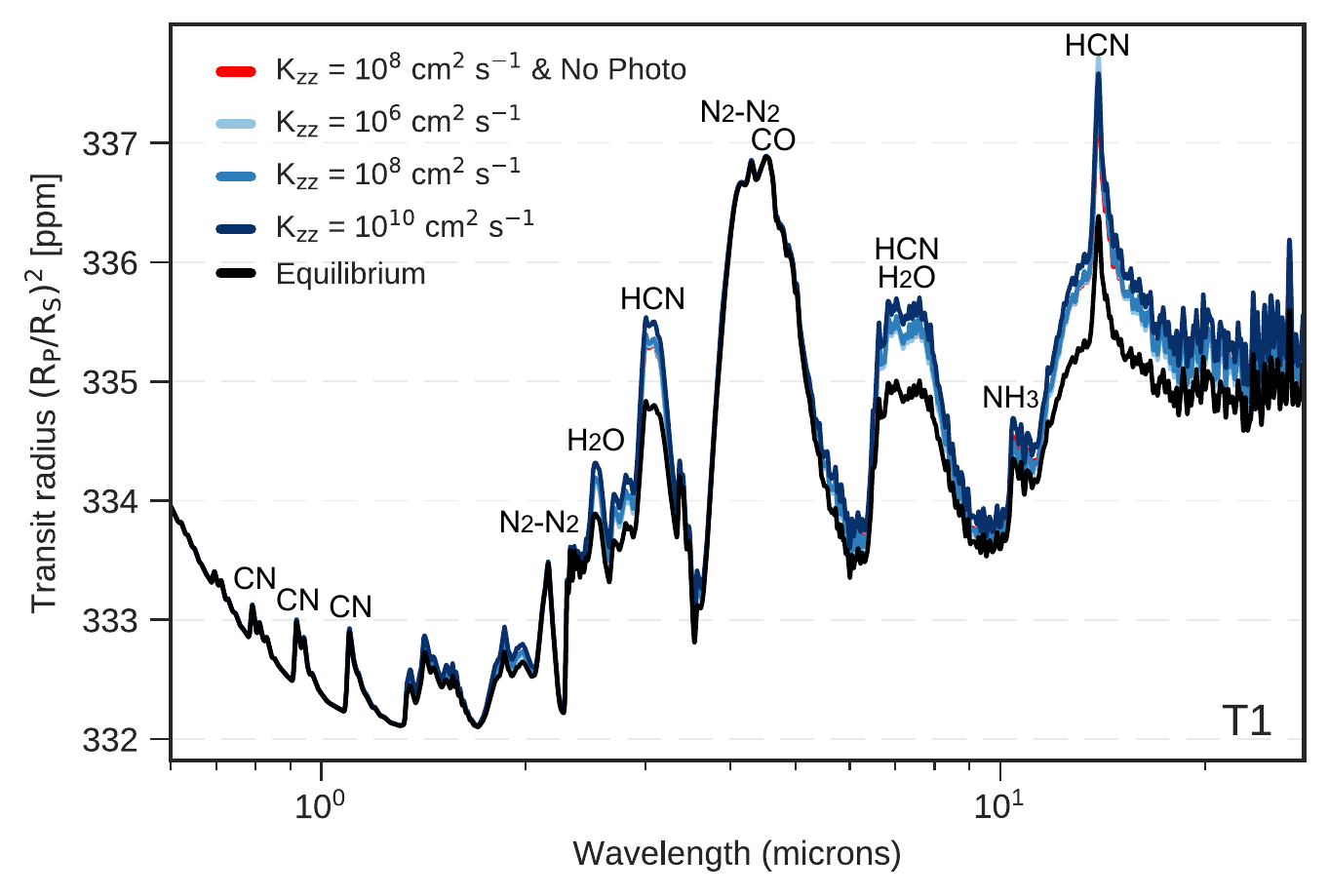}
    \caption{Synthetic transmission spectra for the T1 55 Cnc e profile with different eddy diffusion coefficients. Compositions calculated with C/O = 1.0 and N/O = $3.85 \times 10^{4}$. Spectra in chemical equilibrium (Black) and with no photochemistry (Red) are shown for comparison.}
    \label{fig:transmission_kzz_both_T}
\end{figure}

In Figure \ref{fig:transmission_kzz_both_T} we show the effects of varying eddy diffusion coefficients on the generated spectra. Shown for C/O = 1.0, N/O = $3.85 \times 10^{4}$ and assuming T1, as in the cases studied in Figure \ref{fig:kzz_2709}. Decreasing K\textsubscript{zz} will allow photodissociation of molecules to be more dominant in the upper atmosphere. The vertical mixing will no longer be able to replenish this region with heavier molecular species, resulting in a dramatic reduction of the mean molecular weight, which translates to an increased scale height. However, as shown in Fig. \ref{fig:transmission_kzz_both_T}, this does not result in a significant increase of molecular absorption features. This is because the contributing photospheric region of our models coincides with pressures mainly higher than $10^{-4}$ bar, which are only minimally affected by the varying K\textsubscript{zz} coefficient, thus the slight differences in absorption features. If photochemically produced molecules, such as \ce{CN}, would accumulate in the upper regions of the atmosphere, their absorption features would be significantly influenced by the increased scale height (More on this in Section \ref{sec:discussion_hydrogen_escape}).

\subsubsection{Thermal Emission Spectra}
\begin{figure*}
	\includegraphics[width=\textwidth]{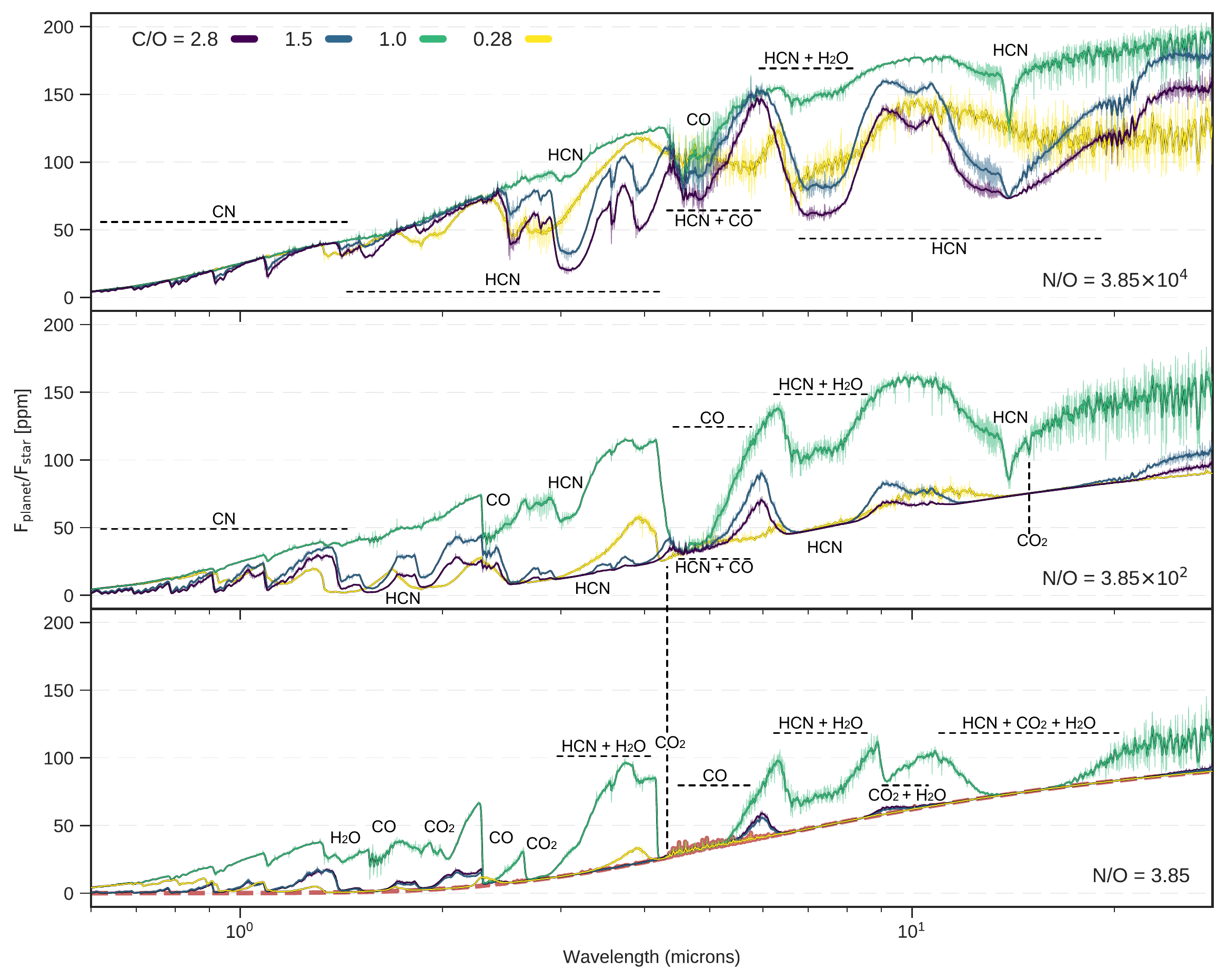}
    \caption{Synthetic emission spectra for the T1 55 Cnc e profile. Shown are the disequilibrium results with a K\textsubscript{zz} = $10^{8}$ cm\textsuperscript{2} s\textsuperscript{-1} for range of C/O and N/O ratios. Each panel represents a different N/O ratio indicated in the lower right corner. Horizontal, dashed lines indicate major contributing species of the broad absorption features. Vertical dashed - narrow features. Carbon-rich atmospheres (Purple, blue) are defined in shape by \ce{HCN} and \ce{CN}. C/O = 1.0 (Green) has a general shape defined by \ce{HCN} and \ce{H2O}. Low C/O ratio spectra (Yellow) are dominated by the absorption of \ce{H2O}, and by \ce{CO2} for the lowest N/O ratios. The orange, dashed curve represents the expected emission from an optically thick atmosphere with an isothermal upper region.}
    \label{fig:2709_e}
\end{figure*}

\begin{figure*}
	\includegraphics[width=\textwidth]{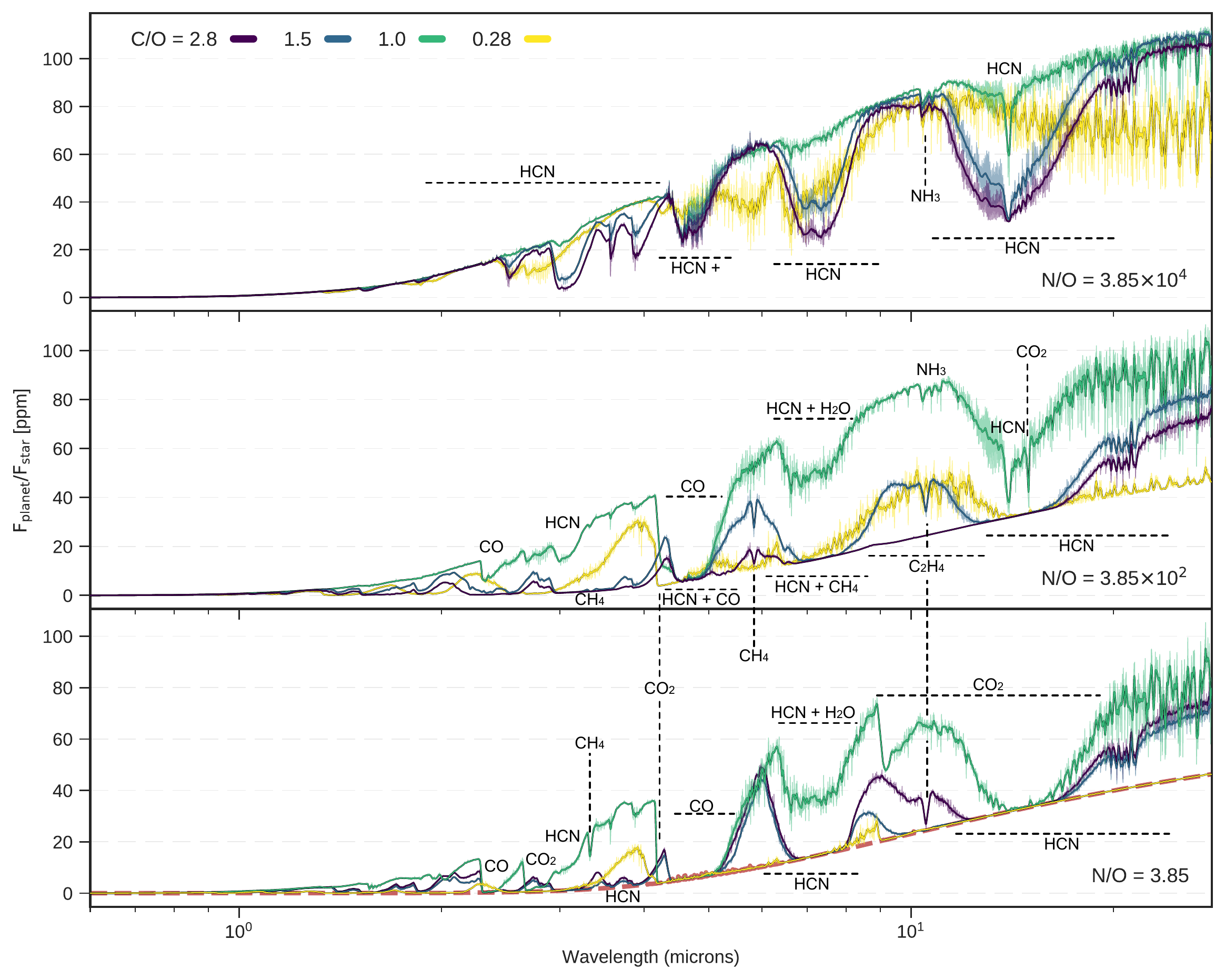}
    \caption{Emission spectra for the T2 55 Cnc e profile with K\textsubscript{zz} = $10^{8}$ cm\textsuperscript{2} s\textsuperscript{-1}. Each panel represents a different N/O ratio indicated in the lower right corner. Horizontal, dashed lines indicate major contributing species of the broad absorption features. Vertical, dashed - narrow features. Carbon-rich atmospheres (Purple, blue) are defined in shape by \ce{HCN} with major \ce{C2H4} absorption for lower N/O ratios. C/O = 1.0 (Green) has a general shape defined by \ce{HCN} and \ce{H2O}. Low C/O ratio spectra (Yellow) are dominated by the absorption of \ce{H2O}, and by \ce{CO2} for the lowest N/O ratios. The orange, dashed curve represents the expected emission from an optically thick atmosphere with an isothermal upper region.}
    \label{fig:1613_e}
\end{figure*}

In thermal emission, the observed molecular features are similar as in transmission. Figures \ref{fig:2709_e} and \ref{fig:1613_e} show emission spectra for the same modelled atmospheres as in transmission.

\textit{C/O > 1.0}: The overall shape is mostly defined by \ce{HCN} absorption. In higher temperatures (T1), \ce{CN} has features up to 1.5 $\micron$. Depending on the N/O ratio, \ce{CO} will be a prominent absorber at 2.3 and 4.6 $\micron$. Both of these molecules show increased absorption with decreased N/O ratio. While we find that \ce{C2H2} is a strong absorber in emission, having similar features to \ce{HCN}, it is never prevalent due to the lack of hydrogen in these atmospheres. \ce{C2H2} would only show up if nitrogen was mostly absent. This, however, may not be completely true, since the available line list for \ce{C2H2} has been compiled only for low temperatures. 

In the T2 spectra (Fig. \ref{fig:1613_e}), \ce{CN} features are no longer visible, but we additionally see \ce{NH3} at 10 $\micron$. As the N/O ratio is reduced \ce{C2H4} becomes an absorber, showing a feature at 10 $\micron$. With the reduced N/O ratio, \ce{CH4} also becomes a strong absorber in colder temperatures.

\textit{C/O = 1.0}: In this case the spectra vary strongly with the N/O ratio. For T1 and high N/O ratios, we see mostly flat spectra with features of \ce{CO}, \ce{HCN} and \ce{H2O}. As the N/O ratio is reduced, the VMR of these species increases, resulting in larger absorption features. \ce{CO2} also starts to contribute at 4.0 and 12 $\micron$. At the lowest N/O ratio considered, \ce{CO2} shows a series of strong absorption features and becomes the major absorber in the 9.0 - 20 $\micron$ region. 

In lower temperatures (T2), we see the same \ce{CO} feature at 4.6 $\micron$ when the N/O ratio is high and additionally at 1.6, 2.3 $\micron$ when the N/O is reduced. \ce{HCN} also has multiple features at 3.0, 7.5 and 11.0 - 15.0 $\micron$. There is also a small \ce{NH3} feature at 10.5 $\micron$.

\textit{C/O < 1.0}: As with transmission, the spectra are dominated by \ce{H2O} and \ce{CO2}. With decreasing N/O ratio, \ce{CO} is increasingly present at 4.6 $\micron$, but is never distinct in these cases.

\begin{figure}
	\includegraphics[width=\columnwidth]{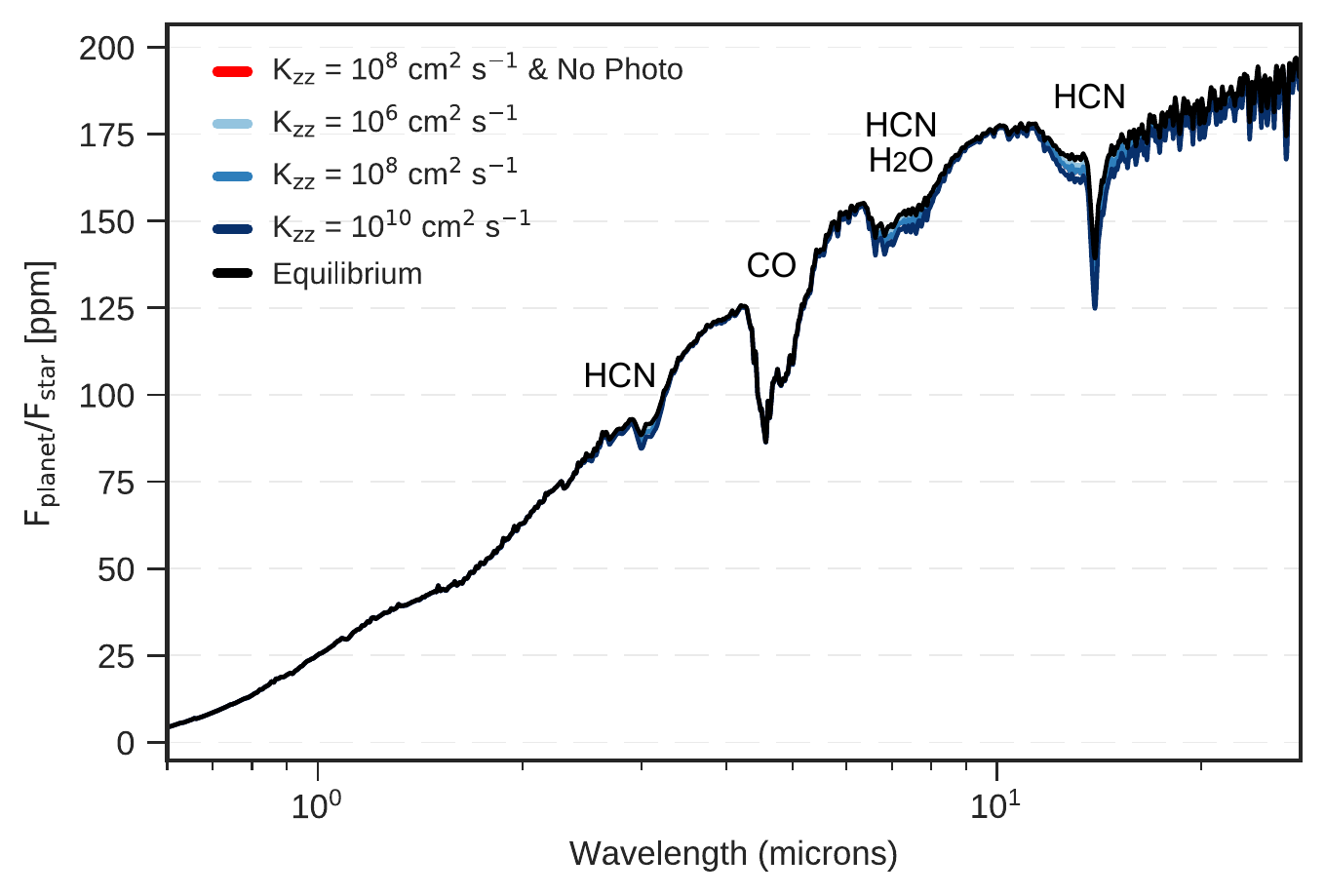}
    \caption{Emission spectra for the T1 55 Cnc e profile with different eddy diffusion coefficients. Compositions calculated with C/O = 1.0 and N/O = $3.85 \times 10^{4}$. Equilibrium (Black) and models without photochemistry (Red) are shown for comparison.}
    \label{fig:emission_kzz_both_T}
\end{figure}

Figure \ref{fig:emission_kzz_both_T} shows the absorption features with varying K\textsubscript{zz}. Just as in transmission, the photospheric pressure coincides with pressures bellow $10^{-4}$ bar, which is not significantly affected by the kinetics. An increase in mixing leads to slightly enhanced absorption of all major molecules, except for \ce{CO}, which is not strongly influenced by the vertical mixing.

\subsubsection{Observability and the JWST}

\begin{figure}
	\includegraphics[width=\columnwidth]{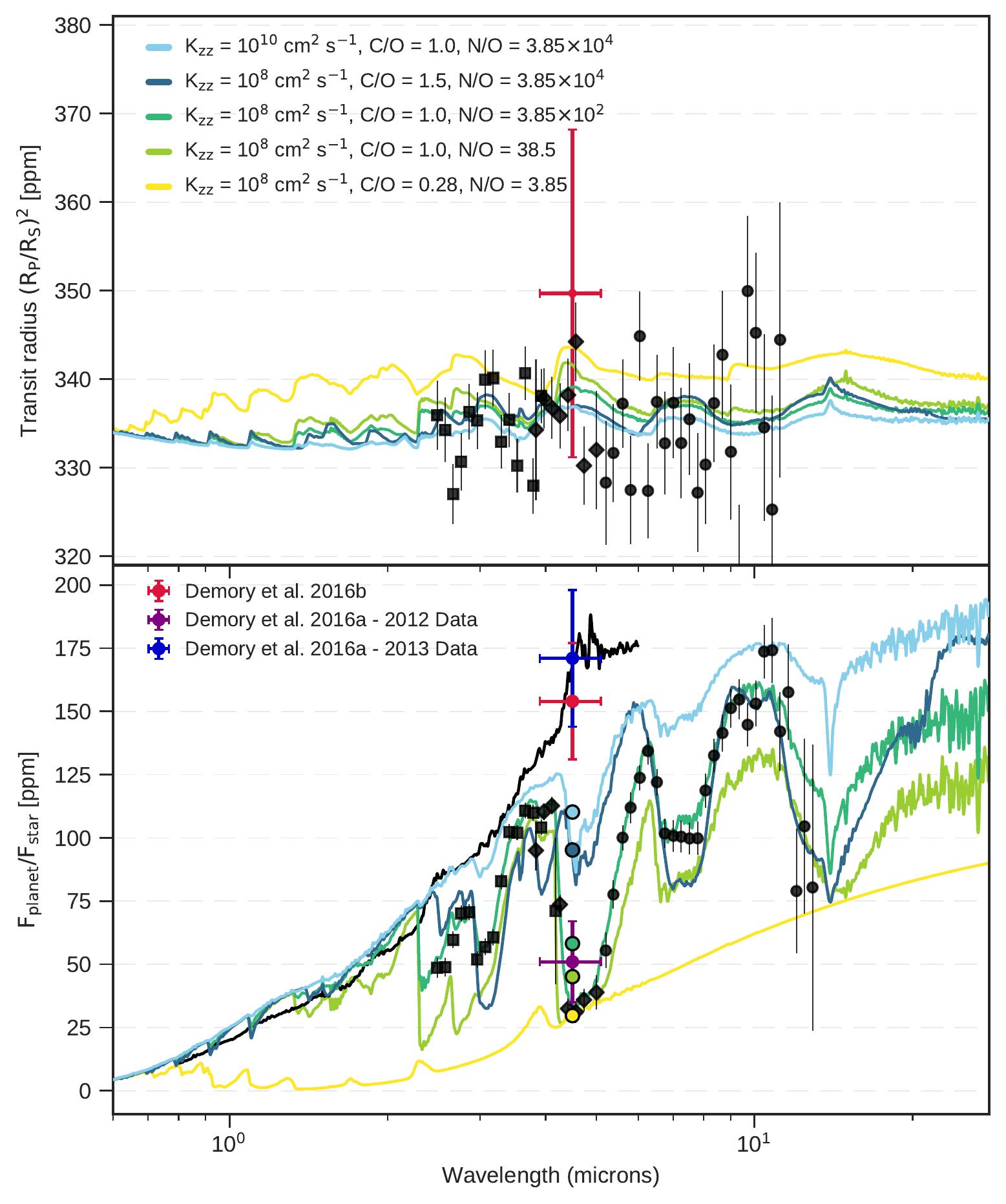}
    \caption{Synthetically generated spectra showing a variation of atmospheric compositions for 55 Cnc e. Transmission is shown in the upper panel, emission in the lower.} Black error bars represent PandExo simulated JWST NIRCam's F322W2 (squares), F444W (diamonds) and MIRI LRS (circles) instrumental noise (1 $\sigma$), modelled with 4 transits or 4 eclipses. For visual clarity, these are binned to R = 15. Shown also are the phase curve observational data from \citet{Demory_2016b,Demory_2016a}. Coloured circles indicate modelled averages. In emission, the black spectrum represents an empty atmosphere while also adapting the mostly empirical 55 Cancri spectrum from \citet{Crossfield_2012}.
    \label{fig:jwst}
\end{figure}

% Intro section
Figure \ref{fig:jwst} shows a selection of modelled atmospheres alongside simulated JWST noise for NIRCam's F322W2, F444W and MIRI LRS instruments, shown for 4 transits or 4 eclipses for each of the instruments. The noise is modelled using the parameters of 55 Cnc e and for visual clarity is binned to a resolution of R = 15. As 55 Cancri is an extremely bright star, it is uncertain whether it would be a suitable target for observations with JWST \citep{Beichman_2014,Nielsen_2016,Batalha_2017}. Thus, the results are not intended to predict the true observability of the super-Earth, but rather depict, in a quantitative manner, the ability of the telescope to detect certain spectral features of different atmospheric compositions around similar, dimmer stars. Keeping in mind that these observations are modelled for a very bright target, the number of transits for a confident feature detection when observing dimmer stars would be increased. Aside from JWST noise simulations, we additionally show phase curve and eclipse observations from \citet{Demory_2016b,Demory_2016a}, captured in the 4.5 $\micron$ IRAC Spitzer band. These are shown for both, transmission and emission.

% Transmission observability and JWST noise
Transmission spectroscopy with JWST for high-mean-molecular-weight super-Earth atmospheres around sun-like stars is not expected to yield high SNR. Since the absorption features of these heavy atmospheres are generally small ($\lesssim$10 ppm), even models with K\textsubscript{zz} = $10^{10}$ cm\textsuperscript{2} s\textsuperscript{-1} are engulfed by the noise of the instruments. For atmospheres with a larger scale height, such as those with low K\textsubscript{zz}, where photodissociation is more efficient, it is feasible, that photochemically produced species, such as \ce{CN}, would have detectable absorption features (> 10 ppm) in the 0.6 - 2.0 $\micron$ region. For \ce{CN}, this is especially true in atmospheres with very low hydrogen content, where it is one of the few species with large opacity values (see Fig. \ref{fig:opacities}) that can efficiently form. While not modelled in this study, highly sensitive JWST's NIRSpec instrument, covering the range of 0.6 - 5.3 $\micron$ may prove to be able to observe transiting planets around stars as bright as 4.5 magnitude in the J band \citep{Ferruit_2014}. Transmission spectroscopy should not be ruled out, as it may be able to detect features of photochemically produced species in high-mean-molecular-weight super-Earth atmospheres.

% Emission observability and JWST noise
Hot super-Earths radiate strongly on their day side. Thermal emission observations should be prioritised for these planets. For C/O > 1.0, \ce{HCN} features between 3.0 to 4.0 $\micron$ are observable with the NIRCam's F322W2 mode. These features are good indicators of high C/O and N/O ratios in N-dominated atmospheres. The $\sim$ 4.0 - 6.0 $\micron$ region, encompassing possible features of \ce{CO2}, \ce{CO}, \ce{H2O}, \ce{HCN}, can also be indicative of the C/O ratio in the atmosphere. This region can be observed with both, NIRCam and NIRSpec instruments. For the MIRI LRS observing mode, the feature at $\sim$7.0 $\micron$ which is due to absorption of both \ce{HCN} and \ce{H2O} can be of importance. While it is hard to distinguish these two molecules with low resolution spectroscopy, the \ce{H2O} contribution is typically broader and less defined. \ce{HCN} has also a very sharp feature at 11 $\micron$, however MIRI's LRS precision drops off significantly at the end of its wavelength coverage due to the star getting fainter in the mid-infrared.

While \ce{CN} has features in our modelled atmospheres, the amplitude of the signal is generally below 10 ppm in emission. This, however, can change if hydrogen is depleted from the atmosphere. In such cases, \ce{CN} signal amplitude can increase to as much as 50 ppm (see discussion in Section \ref{sec:discussion}).

In C/O < 0.5 atmospheres (Yellow curve), we do not see many distinctive features. This is an artefact of our assumption of an isothermal upper atmospheric structure. In reality one would always expect a non-zero gradient for the temperature profile. In Figures \ref{fig:2709_e} and \ref{fig:1613_e}, the dashed curve represents the expected emission from an optically thick atmosphere with an isothermal upper region, which mostly coincides with our low N/O and C/O ratio atmospheres.

% Phase curve observations
We also show phase curve measurements from \citet{Demory_2016b,Demory_2016a}, in both transmission and emission. However, these are not particularly informative in our case, as different data sets are highly variable. While the 2012 data shows strong absorption, which could be caused by either \ce{CO}, \ce{CO2} or \ce{H2O}, the 2013 data indicates that we probe deep into the atmosphere. The overplotted black curve in emission represents an atmosphere with no absorption. For this comparison we use the available measured spectra of 55 Cancri from \citet{Crossfield_2012}, which has higher flux at 5 $\micron$ compared to our generated PHOENIX model. The variability of 55 Cnc e is a known issue, and several possible explanations have been published in the last years, ranging from large scale volcanic plumes, a gas-dust torus, reflective grains or magnetic interaction, to other effects \citep{Demory_2016a,Demory_2016b,Bourrier_2018,2018_Tamburo,2019_Sulis,Folsom_2020}. More observations are needed to decipher the issue.

\section{Discussion}
\label{sec:discussion}

% We present simple models
Our models present NCHO atmospheric cases where nitrogen is the dominant component. While we use chemical kinetics, we do not include other, potentially important processes, such as condensation or surface exchange/atmospheric escape and we also use simplified thermal profiles.

\subsection{C/O \& Graphite Formation}
% C/O in super-Earths and graphite formation
While we do model a large range of C/O ratios in the gas phase, it is not clear whether the atmosphere could sustain a C/O ratio $\gg$ 1.0 in the photosphere (assuming only NCHO gases). The reason being that with carbon saturation, graphite formation begins to take place. Once graphite is a major carbon carrier in the atmosphere, it will begin to deplete the available carbon from the gas-phase chemistry, effectively reducing the C/O ratio. 

\begin{figure}
	\includegraphics[width=\columnwidth]{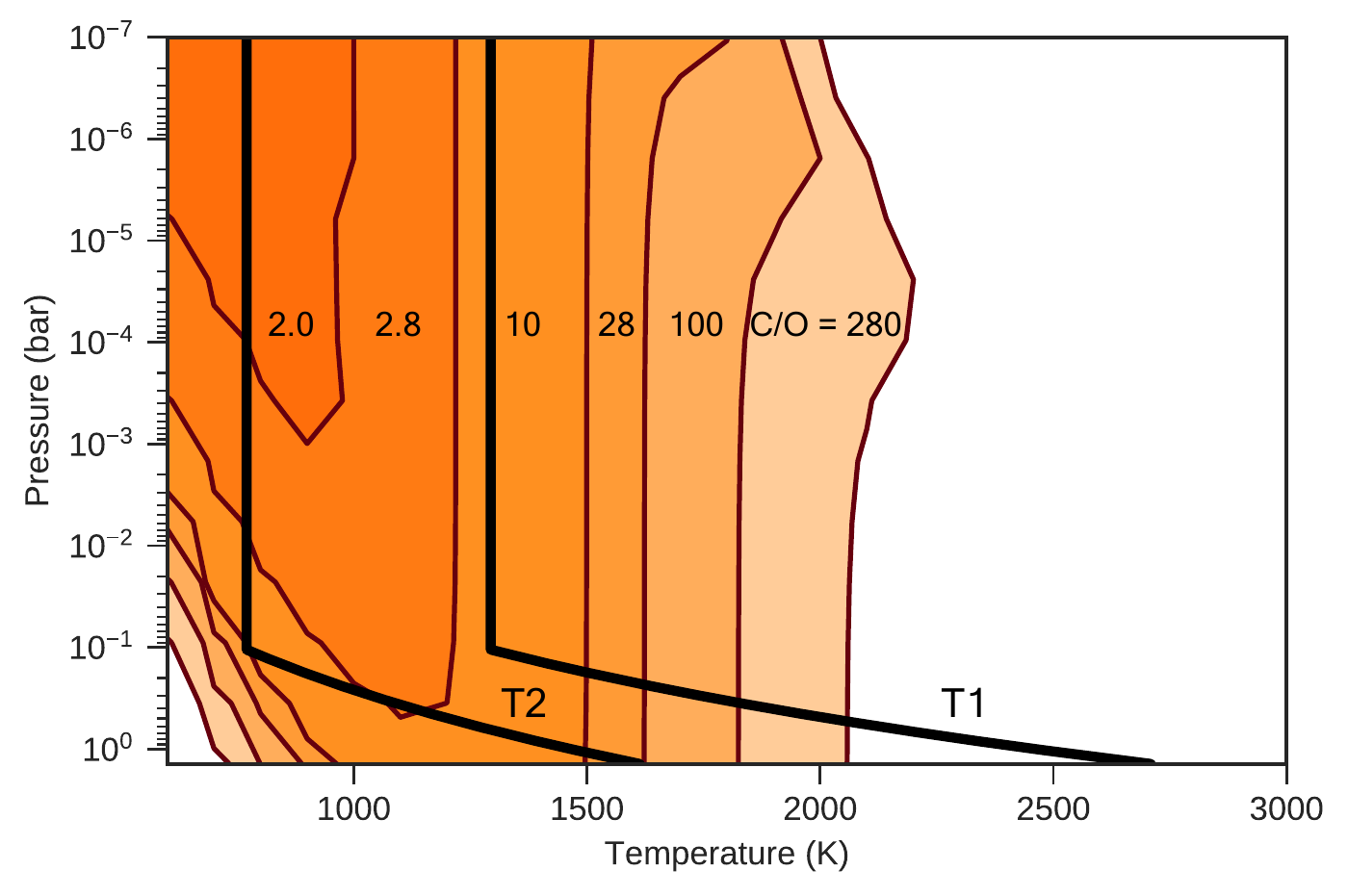}
    \caption{Thermochemical stability of graphite in super-Earth atmospheres. Calculated using Titan's N/O of $3.85 \times 10^{4}$.  The shaded regions indicate where graphite formation is efficient enough for it to be the dominant carbon carrier in the atmosphere. The numbers denote the initial C/O ratio used for the calculation. Overplotted are the explored temperature profiles, T1 and T2. Larger C/O ratio allows graphite formation in a larger pressure range and higher temperatures.}
    \label{fig:graphite}
\end{figure}

We briefly illustrate this in Figure \ref{fig:graphite}, where we use the Chemical Equilibrium with Applications model (CEA2) \citep{1996_Gordon} to model formation of graphite. The shaded regions represent temperature and pressure values where graphite becomes thermochemically the most abundant carbon carrier in the atmosphere. For high C/O ratios, this region is extensive and covers almost the entire range of the temperatures and pressures that we explore in this study. As the C/O drops, the formation efficiency of graphite decreases. Its presence gets confined to lower temperatures and to lower pressures. 
In Fig. \ref{fig:graphite} we show values calculated using Titan's N/O ratio. For this N/O ratio, with T1 profile, we see that atmospheres with C/O = 10 would already have graphite as its major carbon carrier in the upper atmosphere, depleting the carbon in the gas. The C/O ratio would then effectively drop to $\sim$3.0, where graphite would become much less important. Increasing the carbon fraction in our modelled atmospheres will also make graphite formation more favourable, even if the C/O ratio is maintained.

%Graphite formation will also become more favourable if the N/O ratio is reduced in our models, which effectively increases the mass fraction of carbon in the atmosphere (Fig. \ref{fig:graphite_2}). Even at the lowest N/O ratios, USP super-Earths will have sufficiently high temperatures in the photospheric pressures for the C/O ratio above 1.0 to remain stable.

In reality, it is uncertain how the C/O ratio would be affected as graphite formation depends on a multitude of factors besides pressure and temperature. Some of the factors include: metallicity \citep{Lodders_1997,2013b_Moses}; atmospheric kinematics, which could freeze out graphite formation if transportation timescales are shorter than timescales required to remove carbon from its gaseous states \citep{Moses_2013a}; or inclusion of heavier species in our calculations \citep{Lodders_1997}. If carbon managed to condense in the upper atmospheres of hot super-Earth, it would likely slowly settle into the hottest, high-pressure regions and sublimate into the gas-phase. This could result in graphite cloud formation, with the sublimation region being its cloud base. However, it is still uncertain how this would effect the subsequent evolution of the atmosphere, especially when kinetics are taken into account. 

This suggests that atmospheres of hot super-Earths could contain C/O ratios much larger than unity in the photospheric regions and higher above. It is also possible that due to contributions from volcanic activity, oxidation of minerals and reduced gases \citep{Hu_2014,RIMMER_2019}, inclusion of surface exchange would likely show increased C/O ratios for an NCHO atmosphere. 

%\begin{figure}
%	\includegraphics[width=\columnwidth]{Figures/graphite_condense_lab_p.pdf}
%    \caption{\textcolor{red}{(NEW)} Thermochemical stability of graphite in super-Earth atmospheres. Calculated using Titan's N/O of $3.85 \times 10^{4}$.  The shaded regions indicate where graphite formation is efficient enough for it to be the dominant carbon carrier in the atmosphere. The numbers denote the initial C/O ratio used for the calculation. Overplotted are the explored temperature profiles, T1 and T2. Larger C/O ratio allows graphite formation in a larger pressure range and higher temperatures.}
%    \label{fig:graphite_2}
%\end{figure}

\subsection{Impact of Hydrogen Depletion}
\label{sec:discussion_hydrogen_escape}
Another concern is whether USP super-Earths would be able to retain sufficient hydrogen to form detectable levels of volatile species such as \ce{HCN}, \ce{H2O}, \ce{CH4} or unsaturated hydrocarbons. Due to erosion, it is assumed that USP planets would not be able to sustain a large hydrogen or helium envelope over a long period of time \citep{Bourrier_2018}. Failure to detect hydrogen in the exosphere of 55 Cnc e through Ly-$\alpha$ \citep{Ehrenreich_2012} further reinforces this point. On top of this, the close proximity to the star would cause rapid dissociation of most species involving hydrogen, which would allow hydrogen atoms to easily escape the atmosphere. Assuming efficient transport from the lower to the upper regions, hydrogen could be severely, if not completely, depleted over the lifetime of the planet. It is possible that surface emission of reduced gases such as \ce{H2}, \ce{CH4} or \ce{H2S} could resupply the atmosphere with enough hydrogen \citep{Hu_2014}. However, if this would be enough to allow for efficient formation of \ce{HCN} is unknown. 

\begin{figure}
	\includegraphics[width=\columnwidth]{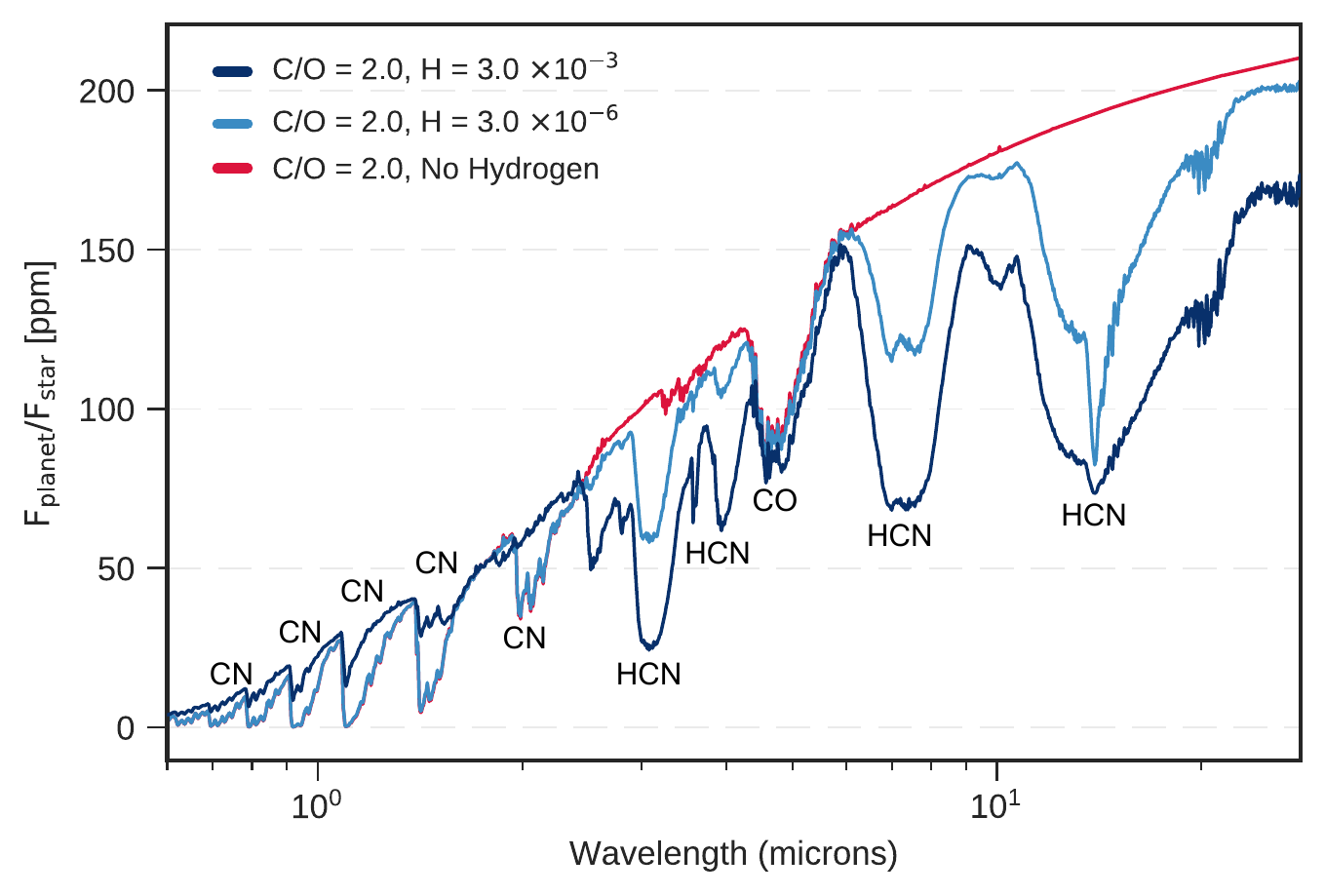}
    \caption{Synthetic spectra showing a variety of \ce{N2} atmospheres where hydrogen is severely depleted. Modelled for the T2 55 Cnc e profile with K\textsubscript{zz} = $10^{8}$ cm\textsuperscript{2} s\textsuperscript{-1} and C/O = 2.0. The indicated hydrogen values are in units of mass fractions.}
    \label{fig:cn}
\end{figure}

In our models we take Titan's elemental composition, containing only 3\% hydrogen, yet a sufficient amount for formation of species like \ce{HCN}. If hydrogen is reduced further, not only will the abundances of \ce{HCN} and hydrocarbons start decreasing, but also of the highly abundant \ce{H2}. The reduction in molecular hydrogen (and other species) would result in much less shielding from UV radiation, allowing photons to penetrate deeper into the atmosphere in the process. This would likely cause a runaway effect, which could leave a significant portion of the atmosphere heavily dissociated, allowing even more hydrogen to escape. And with no hydrogen remaining, our modelled spectra show that it would be dominated by the increased presence of \ce{CN} from 0.6 - 5.0 $\micron$ and \ce{CO} at $\sim$4.6 $\micron$. For high C/O cases, we illustrate this effect in Figure \ref{fig:cn}. With hydrogen fraction of $10 \times 10^{-6}$ and C/O ratio of 2.0, \ce{CN} has features that are close to 40 ppm. We also find that even at such low hydrogen abundance, \ce{HCN} still has multiple absorption features present. While not shown here, for lower C/O ratios, \ce{CO2} and \ce{NO} would also become strong absorbers.

While it is unknown what the hydrogen fraction would be sustained in these atmospheres, a combination of large C/O ratio and depletion of H, could make \ce{CN} an important species for observations.

\subsection{Future Observability of Super-Earths}
% Observations and variability
Our modelled spectra show a variety of observable absorption features in the IR wavelength range applicable to future space missions - JWST and ARIEL. Although our JWST noise levels present an ideal case, where saturation is expected to be the limiting factor, we select the observing modes least affected by it. For JWST, we expect slightly dimmer targets than 55 Cancri to yield adequate measurements. For the brightest targets the upcoming ARIEL telescope \citep{Venot_2018} will prove to be the optimal choice. USP super-Earths: K2-141 b \citep{Malavolta_2018}, Kepler-10 b \citep{Batalha_2011}, WASP-47 e \citep{Becker_2015} and the recently discovered 55 Cancri twin - HD213885 b \citep{Espinoza_2019}, may prove to be ideal for atmospheric characterisation.

In nitrogen dominated atmospheres, carbon-rich and carbon-poor scenarios have vastly different spectral fingerprints. For carbon-rich, \ce{HCN} should be heavily prioritised in observations of hot super-Earths. If planets akin 55 Cnc e did possess a carbon-rich, nitrogen atmosphere, \ce{HCN} would show very clear, distinguishable features in emission, which would be observable with JWST NIRSpec (0.6 - 5.3$\micron$), NIRCam (2.4 - 5.0$\micron$) and with ARIEL (2.0 - 8.0$\micron$). Perhaps MIRI's LRS mode (5.0 - 12.0$\micron$) is less favourable in this situation, mainly due to the fact that the broad \ce{HCN} feature at 7.0 $\micron$ heavily overlaps with other possible molecules, e.g. \ce{C2H2}, \ce{H2O}, even \ce{CO2} (assuming low resolution), and the precision of the instrument quickly falls off at the end of its wavelength range, where \ce{HCN} has another major feature. Although \ce{CN} has weak features in our modelled atmospheres, we find that if hydrogen is severely depleted, its absorption features can manifest as a series of peaks between the range of 0.6 - 5.0 $\micron$ (see Fig. \ref{fig:opacities} and \ref{fig:cn}). Two largest \ce{CN} features would be at around 1.0 - 1.3 $\micron$, reaching close to 40 ppm in emission.

In the study we also explore a large range of N/O ratios. Because hydrogen has a fixed abundance in our models, the reduction in nitrogen results mostly in increased features of all carbon/oxygen species. However, if reduced far enough, \ce{C4H2} becomes very abundant (VMR $\sim 10^{-2}$). Our network is truncated at hydrocarbons containing two carbon atoms, thus this may not represent a realistic value. Rather, it is an indication that instead of \ce{C4H2} various other hydrocarbons would fill the atmosphere. If \ce{C4H2} would be so abundant, it would likely form PAHs and hazes, which we do not explore in this paper. Low N/O ratio would also imply significantly increased abundance of \ce{CO} in hot super-Earth atmospheres, which has a particularly strong feature at $\sim$4.6 $\micron$.

We also note that CIA \ce{N2}-\ce{N2} absorption peaks in the 4.0 - 5.0 $\micron$ region (see Fig. \ref{fig:opacities}), and even though it does not show significant contribution in our atmospheres, we emphasise that the current available \ce{N2}-\ce{N2} CIA data is compiled only for temperatures up to 400 K. Acquisition of high temperature opacity data would allow us to put further constraints on nitrogen dominated atmospheres. 

\section{Conclusion}
Despite observational and modeling efforts \citep{Demory_2016b,Angelo_2017,Hammond_2017,Bourrier_2018,Miguel_2019}, 55 Cnc e remains one of the most mysterious worlds discovered to date. Its large day-night temperature gradient and an eastwards shifted hot spot favours a scenario where it is surrounded by a high-mean-molecular weight atmosphere, possibly composed mainly of \ce{N2}. In this study, we utilise Titan's elemental budget as our starting point and use chemical kinetics to model a range of nitrogen dominated atmospheric compositions, spanning several orders of magnitude in C/O and N/O ratios. From the results we generate synthetic spectra, simulate the noise of several JWST instruments and assess future observability of hot super-Earths. Here are the main conclusions:
\begin{enumerate}

  \item \textbf{In N-dominated hot super-Earth atmospheres, \ce{HCN} is a good indicator of a high C/O ratio, while \ce{H2O} would indicate C/O < 1.0.} These atmospheres vary strongly in composition according to its C/O ratio. \ce{HCN} is found to be very abundant for C/O > 1.0 and is the dominant carbon carrier for C/O > 2.0. \ce{H2O} is only found to be a strong absorber for C/O < 1.0. This behaviour is similar to hot H-dominated atmospheres \citep[See][Figure 18]{Molliere_2015}. In high C/O cases, \ce{CN} and \ce{CO} are also found in high concentrations. At lower temperatures we additionally see \ce{CH4}, \ce{NH3} and \ce{C2H4} amongst the most abundant species. For C/O < 1.0, we find \ce{CO} and \ce{CO2} to be the major carbon-carying absorbers.
  
  \item \textbf{Strong \ce{C2H2} presence instead of \ce{HCN} in the transmission or emission spectrum could indicate that the atmosphere lacks nitrogen.} While \ce{C2H2} abundances are high, it never becomes a species of importance due to strong absorption overlap with \ce{HCN} in our nitrogen atmospheres. The presence of \ce{C2H2} would indicate lower nitrogen abundance and could possibly mean that the background is dominated by hydrogen. We note, however, that the line list available for \ce{C2H2} is low-temperature only.

  \item \textbf{For low N/O ratios we find large \ce{C4H2}, \ce{CO}, and reduced other hydrogen and carbon carrier concentrations}. Due to our reduced chemical network, this might not represent a realistic case, but rather indicate that at low N/O ratios, various hydrocarbons, with two or more carbon atoms, form in the atmosphere. This could also be indicative that graphite clouds, PAHs and hazes are likely to form.
  
  \item \textbf{Detection of \ce{CN} absorption, especially at wavelengths $\geq$ 1.0 $\micron$, could indicate extremely low hydrogen content (and potential hydrogen escape during the evolution of the planet) in nitrogen dominated hot super-Earths with high C/O ratios.} It is expected that due to the intense radiation and temperature, hydrogen would be severely depleted through atmospheric escape. This would make \ce{CN}, \ce{CO} and \ce{NO} notably more abundant. We note that \ce{CN} also has prominent spectral features at wavelengths of 1.0 - 2.0 $\micron$, which, with C/O > 1.0, can reach as much as 40 ppm in emission if hydrogen is sufficiently depleted. Whether such a scenario would be possible is not explored in this paper. 
  
  \item \textbf{\ce{HCN}, \ce{CN} and \ce{CO} should be the prioritised species in observations between 0.6 and 5.0 $\micron$ for hot super-Earths with nitrogen dominated atmospheres.} These species could be detected with the upcoming JWST's NIRSpec, NIRCam observing modes, and with the ARIEL telescope. Thermal emission of USP super-Earths will yield high SNR, with only few occultations needed to put strong constraints on the atmospheric composition. On the other hand, while SNR in transmission spectroscopy is not outstanding in high mean-molecular-weight atmospheres, we find that in low-mixing cases, strong photodissociation results in inflated atmospheres, potentially pushing the absorption signal amplitude of photochemically produced species to observable levels.

\textit{Final remarks:} While we use chemical kinetics to model these atmospheres, we stress the fact that the inclusion of surface exchange, atmospheric escape, condensation/rainout could significantly change the outcome and put further constraints on the elemental ratios. Inclusion of high temperature \ce{N2}-\ce{N2} collision-induced absorption data could potentially allow us to constrain the nitrogen content in future observations of super-Earths.
  
\end{enumerate}
\label{sec:conclusion}

\section*{Acknowledgements}
We thank the anonymous referee for their extensive feedback, which greatly improved the quality of our manuscript. We also thank Alex Cridland, Anna de Graaff, Aur\'{e}lien Wyttenbach, Ignas Snellen, Karan Molaverdikhani, and Laura Kreidberg for their helpful comments and discussion on the topic.

P.M. acknowledges support from the European Research Council under the European Union's Horizon 2020 research and innovation programme under grant agreement No. 832428.
%%%%%%%%%%%%%%%%%%%%%%%%%%%%%%%%%%%%%%%%%%%%%%%%%%



\bibliographystyle{mnras}
\bibliography{paper} % if your bibtex file is called example.bib
%\begin{thebibliography}{99}
%\end{thebibliography}

%%%%%%%%%%%%%%%%%%%%%%%%%%%%%%%%%%%%%%%%%%%%%%%%%%

%%%%%%%%%%%%%%%%% APPENDICES %%%%%%%%%%%%%%%%%%%%%

%\appendix

%\section{Photochemical Pathways}
%\label{photochem_appendix}

%If you want to present additional material which would interrupt the flow of the main paper,it can be placed in an Appendix which appears after the list of references.

%%%%%%%%%%%%%%%%%%%%%%%%%%%%%%%%%%%%%%%%%%%%%%%%%%

% Don't change these lines
\bsp	% typesetting comment
\label{lastpage}
\end{document}